\documentclass[9pt, conference]{IEEEtran}
\IEEEoverridecommandlockouts
\usepackage{cite}
\usepackage{amsmath,amssymb,amsfonts}
\usepackage{amsthm}
\usepackage{algorithmic}
\usepackage{graphicx}
\usepackage{textcomp}
\usepackage{xcolor}
\usepackage{stmaryrd}

\usepackage{multirow}

\usepackage{tikz}
\usepackage{quantikz}
\usepackage{subcaption}

\usepackage{url}

\usepackage{pgfplots}
\usetikzlibrary{patterns}

\newtheorem{theorem}{Theorem}
\newtheorem{example}{Example}
\newtheorem{lemma}{Lemma}

\ifdefined \VersionWithComments
\usepackage{marginnote}
\newcommand{\marginX}{\marginnote{\huge{\quad\quad\textbf{!}\quad\quad}}}
\newcommand{\cyr}[1]{\mbox{}{\color{green!50!black}\marginX{}\textbf{[Yean-Ru}: #1]}}
\newcommand{\lsw}[1]{\mbox{}{\color{orange}\marginX{}\textbf{[Shang-Wei}: #1]}}
\newcommand{\tys}[1]{\mbox{}{\color{blue}\marginX{}\textbf{[YS}: #1]}}
\newcommand{\zhe}[1]{\mbox{}{\color{violet}\marginX{}\textbf{[Zhe}: #1]}}
\newcommand{\david}[1]{\mbox{}{\color{purple}\marginX{}\textbf{[David}: #1]}}
\newcommand{\wzf}[1]{\mbox{}{\color{pink}\marginX{}\textbf{[Tzufan}: #1]}}
\newcommand{\instructions}[1]{{\color{red}\marginX{}\textbf{[Instructions: ``#1'']}}}
\newcommand{\reviewer}[2]{\mbox{}{\color{red}\marginX{}\textbf{[Reviewer #1}: ``#2'']}}
\newcommand{\todo}[1]{\mbox{}{\color{blue}{\marginX{}\textbf{TODO}\ifx#1\\\else:\ \fi #1}}} 
\else
\newcommand{\instructions}[1]{}
\newcommand{\cyr}[1]{}
\newcommand{\lsw}[1]{}
\newcommand{\tys}[1]{}
\newcommand{\zhe}[1]{}
\newcommand{\david}[1]{}
\newcommand{\wzf}[1]{}
\newcommand{\reviewer}[2]{}
\newcommand{\todo}[1]{}
\fi

\def\BibTeX{{\rm B\kern-.05em{\sc i\kern-.025em b}\kern-.08em
    T\kern-.1667em\lower.7ex\hbox{E}\kern-.125emX}}

\begin{document}

\title{A Parallel and Distributed Quantum SAT Solver Based on Entanglement and Quantum Teleportation}

\author{
\IEEEauthorblockN{1\textsuperscript{st} Shang-Wei Lin}
\IEEEauthorblockA{
\textit{Nanyang Technological University}, Singapore \\
shang-wei.lin@ntu.edu.sg
}
\and
\IEEEauthorblockN{2\textsuperscript{nd} Tzu-Fan Wang}
\IEEEauthorblockA{
\textit{National Cheng Kung University}, Taiwan \\
n26114976@gs.ncku.edu.tw
}
\and
\IEEEauthorblockN{3\textsuperscript{rd} Yean-Ru Chen}
\IEEEauthorblockA{
\textit{National Cheng Kung University}, Taiwan \\
chenyr@mail.ncku.edu.tw}
\and
\IEEEauthorblockN{4\textsuperscript{th} Zhe Hou}
\IEEEauthorblockA{
\textit{Griffith University}, Australia \\
z.hou@griffith.edu.au}
\and
\IEEEauthorblockN{5\textsuperscript{th} David Sanán}
\IEEEauthorblockA{
\textit{Singapore Institute of Technology}, Singapore \\
sanan.baena@gmail.com}
\and
\IEEEauthorblockN{6\textsuperscript{th} Yon Shin Teo}
\IEEEauthorblockA{\textit{Continental Automotive}, Singapore \\
yon.shin.teo@continental-corporation.com}
}

\maketitle
\thispagestyle{plain}
\pagestyle{plain}

\begin{abstract}
Boolean satisfiability (SAT) solving is a fundamental problem in computer science. Finding efficient algorithms for SAT solving has broad implications in many areas of computer science and beyond. Quantum SAT solvers have been proposed in the literature based on Grover's algorithm. Although existing quantum SAT solvers can consider all possible inputs at once, they evaluate each clause in the formula one by one sequentially, making the time complexity $O(m)$ --- linear to the number of clauses $m$ --- per Grover iteration. In this work, 
we develop a \emph{parallel} quantum SAT solver, which reduces the time complexity in each iteration from linear time $O(m)$ to constant time $O(1)$ by utilising 
extra entangled
qubits. 
To further improve the scalability of our solution in case of extremely large problems,
we develop a distributed version of the proposed parallel SAT solver based on quantum teleportation such that the total qubits required are shared and distributed among a set of quantum computers (nodes), and the quantum SAT solving is accomplished collaboratively by all the nodes. We have proved the correctness of our approaches and demonstrated them in simulations.
\end{abstract}


\section{Introduction} \label{sec:Intro}



Boolean satisfiability (SAT) solving is a fundamental problem in classical computing. Given a propositional logic formula, SAT determines whether there are truth assignments for propositional variables that make the formula true. SAT has found many applications, including theorem proving, model checking, software and hardware verification, circuit design and optimization, AI planning, scheduling and allocation, etc. Besides these applications, SAT is central in the computation and complexity theories because it is NP-complete, and many other computational problems can be reduced to SAT. Finding efficient algorithms for SAT solving has broad implications for many areas of computer science and beyond.


In classical computing, one of the most widely used algorithms for SAT solving is the DPLL algorithm~\cite{dpll1962}. Its worst-case time complexity is $O(2^n)$, where $n$ is the number of propositional variables in the formula, though in practice it fares better when combined with advanced optimizations, heuristics, parallelization and machine learning. Nonetheless, the exponentially large search space poses a serious challenge for complex problems.

Quantum computing generalizes classical computing from binary bits to quantum bits, which may represent both 0's and 1's simultaneously in a superposition. Another advantage of quantum computers are their innate ability to execute all the possible computational paths simultaneously, known as quantum parallelism. Quantum bits (qubits) can become entangled to each other, a strictly quantum mechanical phenomena with no classical analogue which is also a computing resource that enables quantum computers to achieve \emph{quantum supremacy} over their classical counterparts. These properties make quantum computing very powerful and lead to substantial speed-up compared to certain classical computing algorithms.

\begin{figure}[tb]
\begin{minipage}{\linewidth}
\scalebox{0.5}{
\begin{quantikz}[row sep={7mm,between origins}, column sep=3mm, font=\LARGE]
\lstick{\ket{a}:\ket{+}} & 
[2mm]\gate{X} \gategroup[7,steps=11,style={dotted,fill=yellow!20, inner xsep=5pt, inner ysep=2pt}, background,label style={label position=below,anchor=south,yshift=-0.23cm}]{{\sc $\Omega$}} \gategroup[2,steps=3,style={dashed, rounded corners,fill=blue!20, inner xsep=0pt, inner ysep=0pt}, background,label style={label position=above,anchor=south,yshift=-0.2cm}]{{\sc $C_1$}} & 
[-2mm] \ctrl{1} & [-3mm] \gate{X} & 
[0.5mm] \qw \gategroup[4,steps=3,style={dashed, rounded corners,fill=blue!20, inner xsep=0pt, inner ysep=0pt}, background,label style={label position=above,anchor=south,yshift=-0.2cm}]{{\sc $C_2$}} & [-2mm] \ctrl{2} & [-3mm] \qw & 
[0.5mm] \qw \gategroup[6,steps=3,style={dashed, rounded corners,fill=blue!20, inner xsep=0pt, inner ysep=0pt}, background,label style={label position=above,anchor=south,yshift=-0.2cm}]{{\sc $C_3$}} & [-2mm] \ctrl{4} & [-3mm] \qw & \qw & [-2mm]\qw \gategroup[7,steps=1,style={dashed, rounded corners,fill=cyan!20, inner xsep=0pt, inner ysep=0pt}, background,label style={label position=above,anchor=south,yshift=-0.2cm}]{{\sc $\wedge$}} & [2mm] \qw \gategroup[7,steps=1,style={dashed, rounded corners,fill=red!20, inner xsep=0pt, inner ysep=0pt}, background,label style={label position=above,anchor=south,yshift=-0.2cm}]{{\sc $P$}} & [1mm] \qw \gategroup[7,steps=10,style={dotted,fill=green!15, inner xsep=1pt, inner ysep=1pt}, background,label style={label position=below,anchor=south,yshift=-0.23cm}]{{\sc $\Omega^{-1}$}} & [-2mm] \qw & [-2mm]\ctrl{4} & [-2mm]\qw & [-2mm]\qw & [-3mm]\ctrl{2} & [-2mm]\qw & [-3mm]\gate{X} & [-3mm]\ctrl{1} & [-2mm]\gate{X} & [2mm] \qw \\[1mm]
\lstick{\ket{C_1}:\ket{0}} & \gate{X} & \targ{} & \qw & \qw & \qw & \qw & \qw & \qw & \qw & \qw & \ctrl{5} & \qw & \ctrl{2} & \qw & \qw & \qw & \qw & \qw & \qw & \qw & \targ{} & \gate{X} & \qw \\
\lstick{\ket{b}:\ket{+}} & \qw & \qw & \qw & \gate{X} & \ctrl{1} & \gate{X} & \qw & \qw & \qw & \qw & \qw & \qw & \qw & \qw & \qw & \qw & \gate{X} & \ctrl{1} & \gate{X} & \qw & \qw & \qw & \qw \\[1mm]
\lstick{\ket{C_2}:\ket{0}} & \qw & \qw & \qw & \gate{X} & \targ{} & \qw & \qw & \qw & \qw & \qw & \ctrl{3} & \qw & \ctrl{2} & \qw & \qw & \qw & \qw & \targ{} & \gate{X} & \qw & \qw & \qw & \qw \\
\lstick{\ket{c}:\ket{+}} & \qw & \qw & \qw & \qw & \qw & \qw & \gate{X} & \ctrl{1} & \gate{X} & \qw & \qw & \qw & \qw & \gate{X} & \ctrl{1} & \gate{X} & \qw & \qw & \qw & \qw & \qw & \qw & \qw \\[1mm]
\lstick{\ket{C_3}:\ket{0}} & \qw & \qw & \qw & \qw & \qw & \qw & \gate{X} & \targ{} & \qw & \qw& \ctrl{1} & \qw & \ctrl{1} & \qw & \targ{} & \gate{X} & \qw & \qw & \qw & \qw & \qw & \qw & \qw \\
\lstick{\ket{F}:\ket{0}} & \qw & \qw & \qw & \qw & \qw & \qw & \qw & \qw & \qw & \qw& \targ{} & \gate{Z} & \targ{} & \qw & \qw & \qw & \qw & \qw & \qw & \qw & \qw & \qw & \qw
\end{quantikz}
}
\end{minipage}
\caption{A conventional (sequential) quantum oracle for formula $\mathcal{F}$.} \label{fig:ClassicOracle}
\end{figure}
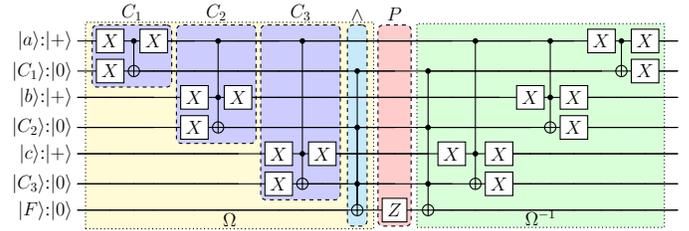

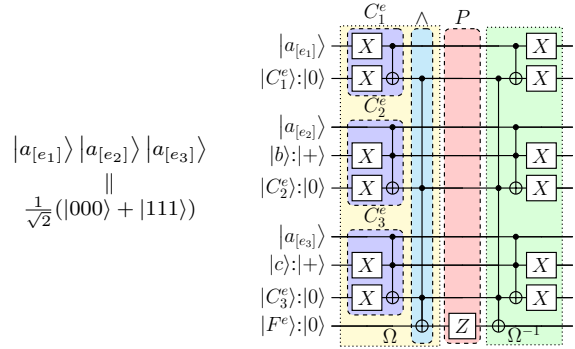
\begin{figure}[tb]
\centering
\begin{minipage}{0.35\linewidth}
\centering
\small
\[
\begin{array}{c}
    \ket{a_{[e_1]}} \ket{a_{[e_2]}} \ket{a_{[e_3]}} \\[1mm]
     \parallel \\
     \frac{1}{\sqrt{2}} (\ket{000} + \ket{111})
\end{array}
\]
\end{minipage}
~
\begin{minipage}{0.6\linewidth}
\scalebox{0.54}{
\begin{quantikz}[row sep={7mm,between origins}, column sep=1mm, font=\LARGE]
\lstick{$\ket{a_{[e_1]}}$} & [4mm]\gate{X} \gategroup[9,steps=3,style={dotted,fill=yellow!20, inner xsep=5pt, inner ysep=2pt}, background,label style={label position=below,anchor=south,yshift=-0.23cm}]{{\sc $\Omega$}} \gategroup[2,steps=2,style={dashed, rounded corners,fill=blue!20, inner xsep=0pt, inner ysep=0pt}, background,label style={label position=above,anchor=south,yshift=-0.2cm}]{{\sc $C_1^e$}} & \ctrl{1} & [3mm] \qw \gategroup[9,steps=1,style={dashed, rounded corners,fill=cyan!20, inner xsep=0pt, inner ysep=0pt}, background,label style={label position=above,anchor=south,yshift=-0.2cm}]{{\sc $\wedge$}} & [4mm] \qw \gategroup[9,steps=1,style={dashed, rounded corners,fill=red!20, inner xsep=0pt, inner ysep=0pt}, background,label style={label position=above,anchor=south,yshift=-0.2cm}]{{\sc $P$}} & [3mm]\qw \gategroup[9,steps=3,style={dotted,fill=green!15, inner xsep=1pt, inner ysep=1pt}, background,label style={label position=below,anchor=south,yshift=-0.23cm}]{{\sc $\Omega^{-1}$}} & \ctrl{1} & \gate{X} & [3mm]\qw \\[1mm]
\lstick{\ket{C_1^e}:\ket{0}} & \gate{X} & \targ{} & \ctrl{3} & \qw & \ctrl{3} & \targ{} & \gate{X} & \qw \\[5mm]
\lstick{$\ket{a_{[e_2]}}$} & \qw \gategroup[3,steps=2,style={dashed, rounded corners,fill=blue!20, inner xsep=0pt, inner ysep=0pt}, background,label style={label position=above,anchor=south,yshift=-2.5mm}]{{\sc $C_2^e$}} & \ctrl{1} & \qw & \qw & \qw & \ctrl{1} & \qw & \qw \\
\lstick{\ket{b}:\ket{+}} & \gate{X} & \ctrl{1} & \qw & \qw & \qw & \ctrl{1} & \gate{X} & \qw \\[1mm]
\lstick{\ket{C_2^e}:\ket{0}} & \gate{X} & \targ{} & \ctrl{4} & \qw & \ctrl{4} & \targ{} & \gate{X} & \qw \\[5mm]
\lstick{$\ket{a_{[e_3]}}$} & \qw \gategroup[3,steps=2,style={dashed, rounded corners,fill=blue!20, inner xsep=0pt, inner ysep=0pt}, background,label style={label position=above,anchor=south,yshift=-2.5mm}]{{\sc $C_3^e$}} & \ctrl{1} & \qw & \qw & \qw & \ctrl{1} & \qw & \qw \\
\lstick{\ket{c}:\ket{+}} & \gate{X} & \ctrl{1} & \qw & \qw & \qw & \ctrl{1} & \gate{X} & \qw \\[1mm]
\lstick{\ket{C_3^e}:\ket{0}} & \gate{X} & \targ{} & \ctrl{1} & \qw & \ctrl{1} & \targ{} & \gate{X} & \qw \\
\lstick{\ket{F^e}:\ket{0}} & \qw & \qw & \targ{} & \gate{Z} & \targ{} & \qw & \qw & \qw
\end{quantikz}
}
\end{minipage}
\caption{A parallel oracle for formula $\mathcal{F}$.} \label{fig:ParallelOracleExample}
\end{figure}

In quantum computing, Grover's algorithm~\cite{grover1996} is able to search for targets in a huge search space with a \emph{quadratic} speed-up compared to classical searching algorithms. Applying it to solve SAT problems has significant theoretical and practical implications. There are two essential components in Grover’s algorithm: (1) an {\em oracle}, and (2) the {\em diffuser}. The oracle answers the “yes/no” question about whether an object in the search space is the target we are looking for. The diffuser tries to maximize the probability of the targets being measured. The detail of Grover’s algorithm is described in Section~\ref{subsec:Grover}. In a nutshell, if one wants to use Grover’s algorithm for a search problem, the key is to provide the oracle. As long as the oracle can correctly identify the targets in the search space, the diffuser, which is standard and independent from the search problem, can help to “extract” the targets. Let us consider the following running example:

\begin{example} \label{ex:RunningExample}
Consider the following Boolean formula $\mathcal{F}$ with three clauses over three Boolean variables.
\vspace{-1mm}
\[
\mathcal{F}: (a) \wedge (\overline{a} \vee b) \wedge (\overline{a} \vee c)
\]
\vspace{-2mm}
$a=1, b=1, c=1$ is the only assignment that makes $\mathcal{F}$ true. \qed
\end{example}

To solve the SAT problem of formula $\mathcal{F}$ by Grover's algorithm, Fernandes et al.~\cite{FS19} proposed an oracle, as shown in Fig.~\ref{fig:ClassicOracle}, where the $C_1$ (cyan) block processes the first clause $(a)$, the $C_2$ block processes the second clause $(\overline{a} \vee b)$, and $C_3$ processes the third clause $(\overline{a} \vee c)$. Even though the three variables $a$, $b$, $c$ are put, respectively, in the $\ket{+}$ superposition state, i.e., $\frac{1}{\sqrt{2}}(\ket{0} + \ket{1})$, to consider all possible inputs at once, the oracle still needs to process each clause one by one sequentially because variable $a$ appears in all the three clauses, and thus the clauses have data dependency. Theoretically, this sequential oracle takes $O(m)$ time complexity, where $m$ is the number of clauses. The readers need not worry about the technical detail here, as it will be briefly introduced in Section~\ref{subsec:QuantumSAT}.
\tys{The bottleneck of the sequential QSAT solver is the decoherence time, i.e. by the time the m-th clause is evaluated, the qubits in the first clause should remain in the same state with high fidelity due to the data dependency. However, the strength of the sequential QSAT is that it does not require highly nonlocal multiqubit GHZ states. In a Quantum computing environment that has a lower decoherence time, a sequential QSAT might be desirable as the maximum number of clauses it can allow in F depends on the decoherence time.}

\tys{On the other hand, for the parallel QSAT, the bottleneck of the number of clauses in F becomes how many maximally entangled qubits can the source provide reliably (depending on the current SOTA and the different platform such as photonics, superconducting qubits, ion traps, it is possible to generate 10~20 maximally entangled qubits in GHZ form). In other words, if the source can only provide 10-qubit GHZ state, then the maximum allowed number of shared variables among the m clauses is 10. Maintaining the entanglement on all of the entangled qubits in the m-qubit GHZ state can be challenging, and the other bottleneck is the precision of each of the local qubit gate on the m-qubits as we require all of them to revert back to the GHZ state after the oracle. If that is not the case, then there is a chance the state will not be 
\vspace{-1mm}
{
\small
\[
\frac{1}{\sqrt{8}} (\ket{\mathbf{\tilde{0}}00} + \ket{\mathbf{\tilde{0}}01} + \ket{\mathbf{\tilde{0}}10} + \ket{\mathbf{\tilde{0}}11} + \ket{\mathbf{\tilde{1}}00} + \ket{\mathbf{\tilde{1}}01} + \ket{\mathbf{\tilde{1}}10} - \ket{\mathbf{\tilde{1}}11})
\]
},but since this is a theoretical paper focusing on quantum algorithms, we should be able to assume that all the gate operations work almost perfectly.
}

In this work, we propose a quantum oracle that processes each clause in parallel, as shown in Fig.~\ref{fig:ParallelOracleExample}, which brings a significant improvement in time complexity from linear time $O(m)$ to constant time $O(1)$. One can observe that the circuit depth in Fig.~\ref{fig:ParallelOracleExample} is much shorter than that in Fig.~\ref{fig:ClassicOracle}, which further reduces the quantum noises during quantum computing. The idea behind our approach is a widely used strategy, ``trade space for time''. We use additional two qubits for variables $a$ so that each clause $C_i$ has its own variable $a_{[e_i]}$ for $i \in \{1,2,3\}$, which makes each clause able to be processed independently in parallel, as the three cyan blocks $C_1^e$, $C_2^e$, $C_3^e$ in Fig.~\ref{fig:ParallelOracleExample}. However, the values of three variables $a_{[e_1]}$, $a_{[e_2]}$, $a_{[e_3]}$ cannot be arbitrary values. They must have the same value as they represent the (single) value of variable $a$ in the formula $\mathcal{F}$. Here comes an interesting question: how do we make sure that the three variables always have the same value? The answer is {\em entanglement}! If we prepare for the three variables the following entangled state
\vspace{-1mm}
\[
\ket{a_{[e_1]}} \ket{a_{[e_2]}} \ket{a_{[e_3]}} = \frac{1}{\sqrt{2}} (\ket{000} + \ket{111}),
\]
then their values will be all $1$ with $\frac{1}{2}$ probability or all $0$ with $\frac{1}{2}$ probability, which captures the exact semantics when solving formula $\mathcal{F}$. The technical details about the proposed parallel oracle and its corresponding diffuser are introduced in Section~\ref{sec:ParallelSAT}. To the best of our knowledge, this is the first work that proposes a parallel quantum SAT solving technique based on entanglement.

The proposed parallel SAT solver gains the improvement in time complexity by paying more (entangled) qubits. 
What if the SAT problem is extremely complex and requires substantial resources?
In such a scenario, {\em distributed quantum computing}~\cite{Grover97,CB97,CEHM99}, adopting the strategy of ``divide and conquer'', emerges as a sub-branch of quantum computing. To overcome this issue of limited resources in a quantum computer, we develop a distributed version of our parallel SAT solver. In this distributed version, the total qubits required are shared and distributed among a set of quantum computers (nodes), and the quantum SAT solving is accomplished collaboratively by all the nodes involved based on quantum teleportation~\cite{BBC93,BPM97, NKL98,RHR04}. The technical detail of the proposed distributed quantum SAT solver is introduced in Section~\ref{sec:DistributedSAT}. To the best of our knowledge, this is also the first work that proposes a distributed quantum SAT solving technique based on quantum teleportation. 


The remaining sections are organized as follows: Section~\ref{sec:Preliminary} reviews necessary technical backgrounds. Section~\ref{sec:ParallelSAT} describes our proposed approach for parallel quantum SAT solving, and Section~\ref{sec:DistributedSAT} extends the parallel approach to a distributed version.
Section~\ref{sec:RelatedWorks} discusses the state-of-the-art approaches to quantum SAT solving and how they relate to this work. Finally, we conclude this work in Section~\ref{sec:Conclusion}.

\section{Preliminaries} \label{sec:Preliminary}


We assume that the readers have basic knowledge in quantum computing, e.g., the {\em tensor product} operation, {\em inner product} operation, {\em outer product} operation, primitive quantum gates (such as $X$, $Z$, $H$, etc), and {\em quantum entanglement}. We use the {\em ket} notation $\ket{\cdot}$ to denote the (column) vector representing the state of a quantum system, and the {\em bra} notation $\bra{\cdot}$ to denote its conjugate transpose. Given two vectors $\ket{v_1}$ and $\ket{v_2}$, we use $\langle v_1 | v_2 \rangle$ to denote their inner product, $|v_1\rangle \langle v_2|$ for their outer product, and $\ket{v_1} \otimes \ket{v_2}$ for their tensor product. For simplicity, we may write $|v_1\rangle \otimes |v_2\rangle$ as $\ket{v_1} \ket{v_2}$, or even $\ket{v_1 v_2}$. When applying an operation on a vector $\ket{v}$, we use $\ket{v'}$ to denote the state of $\ket{v}$ after the operation, or $\ket{v}_t$ to denote the state of $\ket{v}$ at step $t$ during the operation, where $t \in \mathbb{N}$.

\subsection{Grover’s algorithm} \label{subsec:Grover}

Grover’s algorithm~\cite{grover1996} is one of the most well-known quantum algorithms. It is used to solve the search problem for finding target elements in an unsorted database with $N$ elements. 
Due to the characteristic of parallel computation in quantum systems, Grover's algorithm takes $O(\sqrt{N/M})$ operations to find the target element, where $M$ is the number of target elements in the database. It is a quadratic speed up compared with classical methods requiring $O(N)$ operations.
Grover’s algorithm is widely used in many applications, such as cryptography \cite{GL16}, pattern matching \cite{TN22}, etc.

\begin{figure}[tb]
    \centering
    \includegraphics[width=0.95\linewidth]{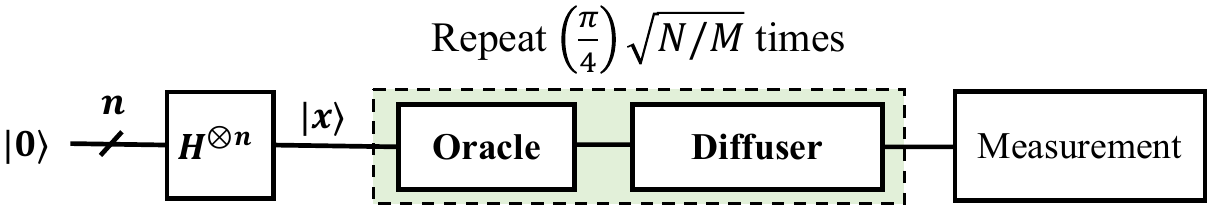}
    \caption{Grover's Algorithm.}
    \label{fig:grover_implement}
\end{figure}

The overall structure of Grover’s algorithm is shown in Fig.~\ref{fig:grover_implement}. The two main operations of it are {\em phase inversion} and {\em inversion about the average}, which are handled by the oracle and diffuser, respectively. 
Initially, the input will be placed in superposition ($|x\rangle$) to evaluate all elements in the database at once. 
Next, the oracle function $U_f$ considers all the possible inputs and marks the target element by applying phase inversion, i.e., $U_f |x\rangle = (-1)^{f(x)} |x\rangle$, in which $f(x) = 1$ for the target element and $f(x) = 0$ for the others.
After the target element is marked, the diffuser applies the {\em inversion about the average} operation, to amplify the probability of the target element, so that one can obtain the result by measurement.
In order to achieve the optimal/maximum probability for the target element to be measured, the two operations (called a Grover iteration) need to be repeated for $(\pi / 4) \sqrt{N/M}$ iterations. The oracle is problem-dependent, while the diffuser is not. Thus, designing the correct oracle is the key to applying Grover’s algorithm. Usually, the number of target elements is unknown before the search, but there are several ways to resolve this issue.
The most common one is to apply quantum counting~\cite{BH98} to obtain the (approximate) number of target elements before using Grover’s algorithm.


\subsection{Conventional Quantum SAT Solving} \label{subsec:QuantumSAT}
Consider the following syntax for SAT formulas in conjunctive normal form (CNF) over a set of Boolean variables $V$:
\[
\begin{array}{lll}
    F & \simeq &  C_1 \wedge C_2 \wedge \cdots \wedge C_m \\
    C & \simeq & l_1 \vee l_2 \vee \cdots \vee l_n \\
    l & \simeq & v \mid \overline{v}
\end{array}
\]

A formula $F$ is a conjunction of $m$ clauses $C_1, C_2, \ldots, C_m$, and each clause $C_i$ is a disjunction of $n$ literals $l_1, l_2, \ldots, l_n$, where $m, n \in \mathbb{N}$. A literal $l_j$ could be a Boolean variable $v$ and called a {\em positive} literal, or the negation of a Boolean variable $\overline v$ and called a {\em negative} literal. We follow the standard semantics in classical logic, i.e., a formula is actually a function $F: \{0,1\}^{|V|} \mapsto \{0, 1\}$ mapping an input vector $\vec{v} \in \{0,1\}^{|V|}$ to true/false (0/1), where $|V|$ denotes the cardinality of $V$. A formula $F$ is \emph{satisfiable} if there exists some $\vec{v} \in \{0,1\}^{|V|}$ such that $F(\vec{v}) = 1$, and we call such $\vec{v}$ a {\em solution} (aka. satisfying assignment) to $F$.  A formula $F$ is \emph{unsatisfiable} if it does not have any solution. 
We do not include Boolean constants true/false in the syntax as they can be rewritten as $(v \vee \overline{v})$ and $(v \wedge \overline{v})$, respectively, and are usually eliminated before SAT solving.


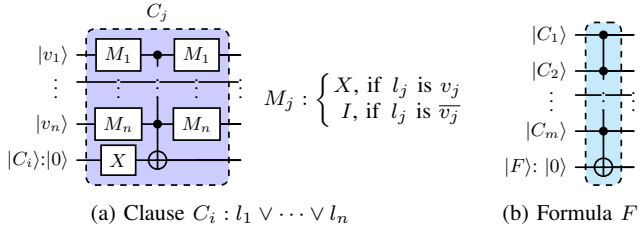
\begin{figure}[tb]
\centering
\begin{subfigure}[b]{0.68\linewidth}
\begin{minipage}{0.54\linewidth}
\scalebox{0.8}{
\begin{quantikz}[row sep={6mm,between origins}, column sep=1mm]
\lstick{\ket{v_1}} & [2mm] \gate{M_1} \gategroup[4,steps=3,style={dashed, rounded corners,fill=blue!20, inner xsep=1pt, inner ysep=1.5pt}, background,label style={label position=above,anchor=south,yshift=-0.2cm}]{{\sc $C_j$}} & \ctrl{1} & \gate{M_1} & [2mm]\qw \\[-1.5mm]
 \hspace{-7mm} \vdots & \vdots & \vdots & \vdots & \vdots \\[1mm]
\lstick{\ket{v_n}} & \gate{M_n} & \ctrl{1} \vqw{-1}  & \gate{M_n} & \qw \qw \\
\lstick{\ket{C_i}:\ket{0}} & \gate{X} & \targ{} & \qw & \qw \qw
\end{quantikz}
}
\end{minipage}
~ 
\begin{minipage}{0.4\linewidth}
\centering
\small
\begin{displaymath}
    M_j: \left\{ \!\!\!
    \begin{array}{l}
      X \mbox{, if } \, l_j \mbox{ is } v_j \\
      \,\, I \mbox{, if } \, l_j \mbox{ is } \overline{v_j}
    \end{array}
    \right.
\end{displaymath}
\end{minipage}
\caption{Clause $C_i: l_1 \vee \cdots \vee l_n$} \label{fig:ClassicClauseCircuit}
\end{subfigure}%
\hfill
\begin{subfigure}[b]{0.27\linewidth}
\centering
\scalebox{0.8}{
\begin{quantikz}[row sep={6mm,between origins}, column sep=2mm]
\lstick{$\ket{C_1}$} & [1mm]\ctrl{1} \gategroup[5,steps=1,style={dashed, rounded corners,fill=cyan!20, inner xsep=1pt, inner ysep=1pt}, background,label style={label position=above,anchor=south,yshift=-0.2cm}]{{\sc }} & [1mm]\qw \\
\lstick{$\ket{C_2}$} & \ctrl{1} & \qw \\[-2mm]
\hspace{-8mm} \vdots & \vdots & \vdots \\
\lstick{$\ket{C_m}$} & \ctrl{1} \vqw{-1} & \qw \\
\lstick{\ket{F}: \ket{0}} & \targ{} & \qw
\end{quantikz}
}
\caption{Formula $F$} \label{fig:ClassicFormulaCircuit}
\end{subfigure}
\caption{The quantum circuit construction scheme for classic oracle.}
\end{figure}

To apply Grover's algorithm for SAT solving of a given formula $F: C_1 \wedge C_2 \wedge \ldots \wedge C_m$, an oracle for $F$ is required. The construction of the quantum circuit for the conventional oracle follows the bottom-up approach\cite{FS19}. The circuit for each clause $C_i$ is constructed first, and then all the clauses are conjuncted together. Fig.~\ref{fig:ClassicClauseCircuit} shows how to construct the circuit for each clause $C_i: l_1 \vee l_2 \vee \ldots \vee l_n$, where the $M_j$ gate depends on literal $l_j$ for $j \in \{1,2, \ldots, n\}$. If $l_j$ is positive, $M_j$ is the $X$ gate, while if $l_j$ is negative, $M_j$ is the $I$ gate. The qubit $\ket{C_i}$ represents the truth value of clause $C_i$.\tys{this is obtained from applying the n-qubit Toffoli gate on the clause $C_i$ that contains n literals.} Once the quantum circuits for all the $m$ clauses are constructed, they are conjuncted by a CNOT gate ($m$-qubit Toffoli gate, to be more precise)\tys{a m-qubit Toffoli gate to be precise} to form the circuit for $F$, as shown in Fig.~\ref{fig:ClassicFormulaCircuit}, where $\ket{F}$ represents the truth value of formula $F$, which is controlled by $\ket{C_i}$ for all $i \in \{1,2, \ldots, m\}$. Fig.~\ref{fig:ClassicOracle} shows the conventional oracle for formula $\mathcal{F}: (a) \wedge (\overline{a} \vee b) \wedge (\overline{a} \vee c)$. The $\Omega$ block is constructed as mentioned previously to identify the solutions of formula $\mathcal{F}$. The $P$ gate is used to give a ``$-1$'' phase to those solutions, and the $\Omega^{-1}$ block is the inverse operation of $\Omega$ to restore each input vector to its initial value for the following diffusion process.
\tys{CNOT gate and its n-qubit extension Toffoli gates are reversible quantum logic gates, so we should be able to obtain the same input state after applying the $\Omega^{-1}$ block.}

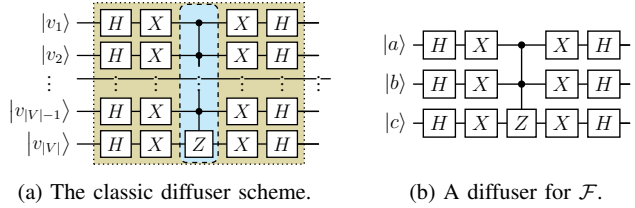
\begin{figure}[tb]
\centering
\begin{subfigure}[b]{0.5\linewidth}
\scalebox{0.62}{
\begin{quantikz}[row sep={5mm,between origins}, column sep=2mm, font=\Large]
\lstick{$\ket{v_1}$} & [3mm]\gate{H} \gategroup[5,steps=5,style={dotted,fill=olive!30, inner xsep=1pt, inner ysep=1pt}, background,label style={label position=above,anchor=south,yshift=-2mm}]{} & \gate{X} & [1mm]\ctrl{2} \gategroup[5,steps=1,style={dashed, rounded corners,fill=cyan!20, inner xsep=0pt, inner ysep=0pt}, background,label style={label position=above,anchor=south,yshift=-0.2cm}]{} & [1mm]\gate{X} & \gate{H} & [2mm]\qw \\[2mm]
 \lstick{$\ket{v_2}$} & \gate{H} & \gate{X} & \ctrl{1} & \gate{X} & \gate{H} & \qw \\
 \hspace{-12mm} \vdots & \vdots & \vdots & \vdots & \vdots & \vdots & \vdots & \\[2mm]
  \lstick{$\ket{v_{|V|-1}}$} & \gate{H} & \gate{X} & \ctrl{1} \vqw{-1} & \gate{X} & \gate{H} & \qw \\[2mm]
  \lstick{$\ket{v_{|V|}}$} & \gate{H} & \gate{X} & \gate{Z} & \gate{X} & \gate{H} & \qw
\end{quantikz}
}
\caption{The classic diffuser scheme.}\label{fig:GeneralClassicDiffuser}
\end{subfigure}
~
\begin{subfigure}[b]{0.45\linewidth}
\centering
\scalebox{0.65}{
\begin{quantikz}[row sep={6mm,between origins}, column sep=2mm, font=\Large]
\lstick{\ket{a}} & \gate{H} & \gate{X} & \ctrl{1} & \gate{X} & \gate{H} & \qw \\[2mm]
\lstick{\ket{b}} & \gate{H} & \gate{X} & \ctrl{1} & \gate{X} & \gate{H} & \qw \\[2mm]
\lstick{\ket{c}} & \gate{H} & \gate{X} & \gate{Z} & \gate{X} & \gate{H} & \qw
\end{quantikz}
}
\\
\vspace{3mm}
\caption{A diffuser for $\mathcal{F}$.}\label{fig:GeneralClassicDiffuserExample}
\end{subfigure}
\caption{Classic Diffuser.} \label{fig:ClassicDiffuser}
\end{figure}

The purpose of the diffuser is to amplify the amplitude of the solution vectors to increase/maximize the probability of them being measured. Fortunately, the diffusion process is independent from the input problems, i.e., different problems can share a general purpose diffuser design. Fig.~\ref{fig:GeneralClassicDiffuser} shows a commonly used diffuser~\cite{grover1996}. The detail of diffusers is omitted here as it is out of the scope. Fig.~\ref{fig:GeneralClassicDiffuserExample} shows the diffuser for three Boolean variables, which works perfectly for formula $\mathcal{F}: (a) \wedge (\overline{a} \vee b) \wedge (\overline{a} \vee c)$.

\section{Parallel Quantum SAT Solver} \label{sec:ParallelSAT}

In this section, we introduce how to parallelize a quantum SAT solver to speed up the SAT solving process. Section~\ref{subsec:ParallelOracle} introduces the proposed parallel oracle 
using entanglement,
and Section~\ref{subsec:ParallelDiffuser} introduces the corresponding parallel diffuser. Discussions and evaluations are then given in Section~\ref{subsec:ParallelSATEvaluation}.

\subsection{Parallel Oracle} \label{subsec:ParallelOracle}


Let $V$ be a set of Boolean variables and $F: C_1 \wedge C_2 \wedge \cdots \wedge C_m$ be a Boolean CNF formula over $V$ with $m$ clauses, where $m \in \mathbb{N}$. If a variable $v \in V$ is shared by $k$ clauses in $F$ where $k \in \mathbb{N}$, we call it a {\em shared} variable. For formula $F$, we define its {\em expanded} formula with respect to $v$, denoted by $F^e_{v}$, obtained by replacing each occurrence of variable $v$ with a \emph{(fresh) expanded} variable $v_{[e_i]}$ where $i \in \{1,2, \ldots, k\}$ and $v_{[e_1]} = v$. Since $v_{[e_1]} = v$, we may use these two symbols interchangeably, and we use $\llbracket v \rrbracket$ to denote the set of expanded variables $\{ v, v_{[e_2]}, \ldots, v_{[e_k]} \}$. We generalize the definition of expanded formulas to the whole set $V$, and the expanded formula is denoted by $F_{V}^e$ or even $F^e$, in which \emph{every} shared variable is treated in the above manner. We use $V^e = \bigcup_{v \in V} \llbracket v \rrbracket$ to denote the set of Boolean variables of $F_{V}^e$, and each clause in $F_{V}^e$ is denoted by $C_j^e$ where $j \in \{1,2, \ldots, m\}$. Example~\ref{ex:ExpandedFormula} illustrates our definitions.

\begin{example} \label{ex:ExpandedFormula}
Consider formula $\mathcal{F}$ over $V=\{a, b, c\}$ in Example~\ref{ex:RunningExample}. The variable $a$ appears in three clauses, so we can obtain the following expanded formula, where $a = a_{[e_1]}$:
\vspace{-1mm}
\[
\mathcal{F}_a^e: (a_{[e_1]}) \wedge (\overline{a_{[e_2]}} \vee b) \wedge (\overline{a_{[e_3]}} \vee c)
\]
As $a$ is the only shared variable, the expanded formula $F_V^e$ would be $F_a^e$, where $C_1^e = (a_{[e_1]})$, $C_2^e = (\overline{a_{[e_2]}} \vee b)$, $C_3^e = (\overline{a_{[e_3]}} \vee c)$, and $V^e = \{a_{[e_1]}, a_{[e_2]}, a_{[e_3]}, b, c \}$.
\qed
\end{example}

It is obvious that a Boolean formula $F$ may not be logically equivalent to its expanded formula $F_V^e$. However, if $F_V^e$ is {\em equivalently expanded}, i.e., it satisfies the following condition:
\vspace{-1mm}
\[
v_{[e_1]} \Leftrightarrow v_{[e_2]} \Leftrightarrow \cdots \Leftrightarrow v_{[e_k]} \mbox{ for all } v \in V
\]
then an input vector $\vec{v} \in \{0,1\}^{|V|}$ for formula $F$ uniquely determines an input vector $\vec{v^e} \in \{0,1\}^{|V^e|}$  for formula $F_V^e$, and vice versa. In such cases, Lemma~\ref{lm:ExpandedFormula} proves that $\vec{v}$ is a solution to $F$ if and only if $\vec{v^e}$ is a solution to $F_V^e$. Let us consider $\mathcal{F}$ in Example~\ref{ex:RunningExample} again. If $a_{[e_1]} \Leftrightarrow a_{[e_2]} \Leftrightarrow a_{[e_3]}$, then $\mathcal{F}_V^e \Leftrightarrow \mathcal{F}$. 

\begin{lemma} \label{lm:ExpandedFormula}
Given a formula $F$ over $V$, if $F$ is equivalently expanded to $F_V^e$, then $\vec{v}$ is a solution to $F$ iff $\vec{v^e}$ is a solution to $F_V^e$.
\end{lemma}

\begin{proof}




Consider a shared variable $v$ with $k$ expanded variables $\{v_{[e_1]}, v_{[e_2]}, \ldots, v_{[e_k]}\}$. Since $v$ is logically equivalent to every of its expanded variable $v_{[e_i]}$ where $i \in \{1,2, \ldots, k\}$ (note that $v = v_{[e_1]}$), the value of $v$ in the solution $\vec{v}$ must be the same as the value of each of $\{v_{[e_1]}, v_{[e_2]}, \ldots, v_{[e_k]}\}$ in the solution $\vec{v^e}$. As a result, we can substitute that value into the formulae $F$ and $F_V^e$ and unify all the expanded variables of $v$. Performing the same for every (shared) variable and substituting the values into the formula, the two formulae $F$ and $F_V^e$ become syntactically identical after all the value substitutions. Therefore, $\vec{v}$ makes $F$ true iff $\vec{v^e}$ makes $F_V^e$ true.
\end{proof}

Based on Lemma~\ref{lm:ExpandedFormula}, given a CNF formula $F$ over $V$, our parallel oracle operates on its equivalently expanded formula $F^e$. 
But how can we ensure that those expanded variables are logically equivalent? The answer is {\em entanglement}! That is, for each variable $v \in V$ shared among $k$ clauses, we prepare the following entangled state initially for $v$ and its expanded variables:
\vspace{-1mm}
\[
\ket{v_{[e_1]}} \ket{v_{[e_2]}} \cdots \ket{v_{[e_k]}} = \frac{1}{\sqrt{2}} (\ket{0}^{\otimes_k} + \ket{1}^{\otimes_k})
\]
In this setting, each shared variable and its expanded variables will be all $\ket{0}$ with $\frac{1}{2}$ probability or be all $\ket{1}$ with $\frac{1}{2}$ probability.

\begin{figure}[tb]
\centering
\begin{subfigure}[b]{0.4\linewidth}
\centering
\scalebox{0.75}{
\begin{quantikz}[row sep={6mm,between origins}, column sep=2mm]
\lstick{\ket{v_1}} & [1mm] \gate{M_1} \gategroup[5,steps=3,style={dashed, rounded corners,fill=blue!20, inner xsep=1pt, inner ysep=1.5pt}, background,label style={label position=above,anchor=south,yshift=-0.2cm}]{{\sc $C_i^e$}} & \phase[red]{\overline{l_1}} & [-2mm]\ctrl{1} & [1mm]\qw \rstick{\ket{v_1'}} \\
\lstick{\ket{v_2}} & \gate{M_2} & \phase[red]{\overline{l_2}} & \ctrl{1} & \qw \qw \rstick{\ket{v_2'}} \\
 \hspace{-7mm} \vdots & \vdots & \vdots & \vdots & \vdots \\[2mm]
\lstick{\ket{v_n}} & \gate{M_n} & \phase[red]{\overline{l_n}} & \ctrl{1} \vqw{-1} & \qw \qw \rstick{\ket{v_n'}} \\
\lstick{\ket{C_i^e}: \ket{0}} & \gate{X} & \qw & \targ{} & \qw \qw \rstick{\ket{{C_i^e}'}}
\end{quantikz}
}
\caption{Clause $C_i^e$} \label{fig:ClauseCircuit}
\end{subfigure}%
\hfill
\begin{subfigure}[b]{0.28\linewidth}
\centering
\scalebox{0.75}{
\begin{quantikz}[row sep={6mm,between origins}, column sep=2mm]
\lstick{$\ket{C_1^e}$} & [1mm]\ctrl{1} \gategroup[5,steps=1,style={dashed, rounded corners,fill=cyan!20, inner xsep=1pt, inner ysep=1pt}, background,label style={label position=above,anchor=south,yshift=-0.2cm}]{{\sc }} & [1mm]\qw \\
\lstick{$\ket{C_2^e}$} & \ctrl{1} & \qw \\
\hspace{-8mm} \vdots & \vdots & \vdots \\
\lstick{$\ket{C_m^e}$} & \ctrl{1} \vqw{-1} & \qw \\
\lstick{\ket{F^e}: \ket{0}} & \targ{} & \qw \rstick{\ket{{F^e}'}}
\end{quantikz}
}
\\ 
\vspace{2mm}
\caption{Formula $F^e$} \label{fig:FormulaCircuit}
\end{subfigure}%
\hfill
\begin{subfigure}[b]{0.3\linewidth}
\centering
\scalebox{0.75}{
\begin{quantikz}[row sep={6mm,between origins}, column sep=2mm]
\lstick{\ket{v_1}} & [1mm] \ctrl{1} \gategroup[5,steps=2,style={dashed, rounded corners,fill=gray!20, inner xsep=1pt, inner ysep=1.5pt}, background,label style={label position=above,anchor=south,yshift=-0.2cm}]{{\sc ${C_i^e}^{-1}$}} & \gate{M_1} & [1mm]\qw \rstick{\ket{v_1'}} \\
\lstick{\ket{v_2}} & \ctrl{1} & \gate{M_2} & \qw \qw \rstick{\ket{v_2'}} \\
 \hspace{-7mm} \vdots & \vdots & \vdots & \vdots \\[2mm]
\lstick{\ket{v_n}} & \ctrl{1} \vqw{-1} & \gate{M_n} & \qw \qw \rstick{\ket{v_n'}} \\
\lstick{\ket{C_i^e}} & \targ{} & \gate{X} & \qw \qw \rstick{\ket{{C_i^e}'}}
\end{quantikz}
}
\caption{Inverse Circuit} \label{fig:ClauseInverseCircuit}
\end{subfigure}%
\caption{Quantum circuit construction scheme for clauses and formula.}
\end{figure}
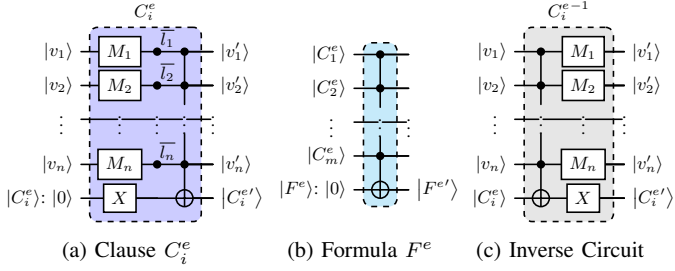

The proposed parallel oracle construction is a bottom-up approach. Suppose the expanded formula is $F_V^e: C_1^e \wedge C_2^e \wedge \cdots \wedge C_m^e$. The quantum circuit of each clause $C_i^e$ is constructed first for all $i \in \{1,2,\ldots, m \}$, and all the $m$ clause circuits are then conjuncted. Fig.~\ref{fig:ClauseCircuit} shows how to construct the circuit for each clause $C_i^e: l_1 \vee l_2 \vee \ldots \vee l_n$, where the qubit $\ket{C_i^e}$ represents the truth value (initially $\ket{0}$) of clause $C_i^e$. Notice that the $M_j$ gate here depends on literal $l_j$ for $j \in \{1,2,\ldots,n\}$, exactly the same as in Fig.~\ref{fig:ClassicClauseCircuit}, i.e., if $l_j$ is negative, $M_j$ would be the $I$ gate ; otherwise, $M_j$ would be the $X$ gate. Lemma~\ref{lem:SAT-Clause} proves the correctness of the clause construction. 


\begin{lemma}[Clause Correctness] \label{lem:SAT-Clause}
$\ket{{C_i^e}'} = \ket{1}$ $\Leftrightarrow$ clause $C_i^e$ is true. 
\end{lemma}

\begin{proof}
Given a clause $C_i^e = l_1 \vee l_2 \vee \cdots \vee l_n$, if $l_j$ is $\overline{v_j}$ where $j \in \{1,2, \ldots, n\}$, then $M_j$ in Fig.~\ref{fig:ClauseCircuit} would be the $I$ gate; otherwise, $M_j$ would be the $X$ gate. Thus, we have $M_j(\ket{v_j}) = \overline{l_j}$ in all cases, as the red notations in Fig.~\ref{fig:ClauseCircuit}.

Since $\ket{{C_i^e}} = \ket{0}$, $X(\ket{{C_i^e}}) = \ket{1}$. If $\ket{{C_i^e}'} = \ket{0}$, by the property of the CNOT gate ($n$-qubit Toffoli gate), every $\overline{l_i}$ must be $1$, so we have $\ket{{C_i^e}'} = \ket{0} \Leftrightarrow (\overline{l_1} = 1) \wedge \cdots \wedge (\overline{l_n} = 1)$. If we apply negation on both sides, we have
$\ket{{C_i^e}'} = 1 \Leftrightarrow \overline{(\overline{l_1} = 1)} \vee \cdots \vee \overline{(\overline{l_n} = 1)}$. Thus, 
$\ket{{C_i^e}'} \!= \!1 \Leftrightarrow (l_1 = 1) \vee \cdots \vee (l_n = 1)$, and the right hand side means that the clause $C_i^e$ is true.
\end{proof}

Once all the $m$ clauses are constructed, they are conjuncted by a $m$-qubit Toffoli gate, as shown in Fig.~\ref{fig:FormulaCircuit}, in which $\ket{F^e}$ is the qubit (initially 0) representing the truth value of formula $F^e$ controlled by the $m$ qubits $\ket{C_i^e}$ for $i \in \{ 1,2,\ldots, m \}$. Lemma~\ref{lem:SAT-Formula} proves the correctness of the formula construction.

\begin{lemma}[Formula Correctness]\label{lem:SAT-Formula}
$\ket{{F^e}'} \!=\! \ket{1}$ $\Leftrightarrow$ formula $F^e$ is true.
\end{lemma}

\begin{proof}
$\ket{{F^e}'} = \ket{1} \Leftrightarrow \ket{C_i^e} = \ket{1}$ for all $i \in \{1,2, \ldots, m\}$. Based on Lemma~\ref{lem:SAT-Clause}, $\ket{C_i^e} = \ket{1} \Leftrightarrow$ clause $C_i^e$ is true. Thus, we can conclude that $\ket{{F^e}'} = \ket{1}$ $\Leftrightarrow$ formula $F^e$ is true.
\end{proof}

\begin{figure}[tb]
\centering
\scalebox{0.66}{
\begin{quantikz}[row sep={4mm,between origins}, column sep=4mm, font=\Large]
 & \qw \qwbundle{} & \gate[wires=2, style={fill=blue!20}][11mm]{C_1^e} \gategroup[6,steps=2,style={dotted,fill=yellow!20, inner xsep=5pt, inner ysep=2pt}, background,label style={label position=below,anchor=south,yshift=-0.23cm}]{{\sc $\Omega$}} & \qw \gategroup[6,steps=1,style={dashed, rounded corners,fill=cyan!20, inner xsep=0pt, inner ysep=0pt}, background,label style={label position=above,anchor=south,yshift=-0.2cm}]{{\sc $\wedge$}} & [-2mm]\qw \slice{1} & [6mm]\qw \gategroup[6,steps=1,style={dashed, rounded corners,fill=red!20, inner xsep=0pt, inner ysep=0pt}, background,label style={label position=above,anchor=south,yshift=-0.2cm}]{{\sc $P$}} & [-5mm] \qw \slice{2} & [9mm] \qw \gategroup[6,steps=2,style={dotted,fill=green!20, inner xsep=5pt, inner ysep=2pt}, background,label style={label position=below,anchor=south,yshift=-0.23cm}]{{\sc $\Omega^{-1}$}} & [-1mm]\gate[wires=2, style={fill=gray!20}]{C_1^{e^{-1}}} & \qw \slice{3} & [2mm]\qw \\
 \lstick{\ket{C_1^e}} & \qw & \qw & \ctrl{1} & \qw & \qw & \qw & \ctrl{1} & \qw & \qw & \qw \\[2mm]
  \hspace{-10mm} \vdots & \hspace{-5mm} \vdots & \vdots & \vdots & & \vdots & & \vdots & \hspace{-2mm} \vdots & & \vdots \\[4mm]
 & \qw \qwbundle{} & \gate[wires=2, style={fill=blue!20}][11mm]{C_m^e} & \qw & \qw & \qw & \qw & \qw &[-1mm]\gate[wires=2, style={fill=gray!20}]{C_m^{e^{-1}}} & \qw & \qw \\
 \lstick{\ket{C_m^e}} & \qw & \qw & \ctrl{1} \vqw{-2} & \qw & \qw & \qw & \ctrl{1} \vqw{-2} & \qw & \qw & \qw \\[2mm]
 \lstick{\ket{F^e}: \ket{0}} & \qw & \qw & \targ{} & \qw & \gate{Z} & \qw & \targ{} & \qw & \qw & \qw
\end{quantikz}
}
\caption{The parallel oracle construction scheme.}\label{fig:ParallelOracle}
\end{figure}
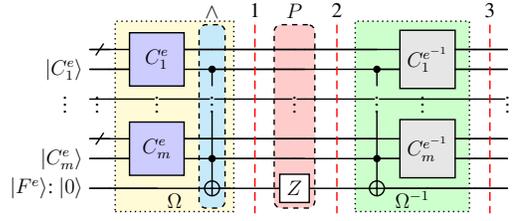

Fig.~\ref{fig:ParallelOracle} shows the quantum circuit construction for the whole parallel oracle $\mathcal{O} = \Omega^{-1}( P (\Omega))$, where the $\Omega$ block is constructed by composing the building blocks of clause circuits and their conjunction; the $P$ block applies a $Z$ gate on the $\ket{F^e}$ qubit to give a ``$-1$'' phase to the input vector when $\ket{F^e}$ is $\ket{1}$, i.e., when formula $F^e$ evaluates to true; the $\Omega^{-1}$ block is the inverse operation of $\Omega$ to restore the input vector back to its initial value for the following diffusion process. Notice that the $C_i^{e^{-1}}$ circuit is the inverse operation of $C_i^e$. Its construction is shown in Fig.~\ref{fig:ClauseInverseCircuit}. The correctness of the proposed parallel oracle $\mathcal{O}$ is proved in Theorem~\ref{thm:ParallelOracleCorrectness}.

\begin{theorem}[Parallel Oracle Correctness] \label{thm:ParallelOracleCorrectness}
Let $\vec{v^e}$ be the input vector of formula $F^e$. Our parallel oracle $\mathcal{O}$ ensures the following:
\begin{displaymath}
    \mathcal{O}(\ket{\vec{v^e}}) = \left\{ \!\!\!
    \begin{array}{rcl}
      \ket{\vec{v^e}} & \mbox{, if } & F^e(\vec{v^e}) = 0 \\[1mm]
      - \ket{\vec{v^e}} & \mbox{, if } & F^e(\vec{v^e}) = 1
    \end{array}
    \right.
\end{displaymath}
\end{theorem}

\begin{proof}
Let $\ket{s}: \ket{\vec{v^e}} \ket{F^e}$ be the state of $\mathcal{O}$, and $\ket{s}_t$ denotes the state of $\mathcal{O}$ at step~$t$, highlighted as dotted red line in Fig.~\ref{fig:ParallelOracle}. Initially, $\ket{s}_0 = \ket{\vec{v^e}}_0 \ket{0}$. At Step~$1$, state $\ket{s}_1$ could be $\ket{\vec{v^e}}_1 \ket{0}$ or $\ket{\vec{v^e}}_1 \ket{1}$, as $\ket{F^e}$ could be either $\ket{0}$ or $\ket{1}$. At Step~$2$, a $Z$ gate is applied on $\ket{F^e}$. Since $Z(\ket{0}) = \ket{0}$ and $Z(\ket{1}) = -\ket{1}$, state $\ket{s}_2$ would be either $\ket{\vec{v^e}}_1 \ket{0}$ or $-\ket{\vec{v^e}}_1 \ket{1}$. Based on Lemma~\ref{lem:SAT-Formula}, $\ket{F^e} = 1 \Leftrightarrow$ formula $F^e$ is true $\Leftrightarrow$ $\vec{v^e}$ is a solution to $F^e$. Thus, if state $\ket{s}_t$ has a ``$-1$'' phase when $t \geq 2$, then $\ket{\vec{v^e}}_0$ in that state is a solution to $F^e$. At step~$3$, the $\Omega^{-1}$ block is applied to restore $\ket{\vec{v^e}}$ and $\ket{F^e}$ back to their initial values. Thus, $\ket{s}_3$ is either $\ket{\vec{v^e}}_0 \ket{0}$ or $-\ket{\vec{v^e}}_0 \ket{0}$. The former is the case: $\mathcal{O}(\ket{\vec{v^e}}) = \ket{\vec{v^e}}$ when $F^e(\vec{v^e}) = 0$, while the latter is the case: $\mathcal{O}(\ket{\vec{v^e}}) = -\ket{\vec{v^e}}$ when $F^e(\vec{v^e}) = 1$.
\end{proof}

Let us get back to our running example $\mathcal{F}: (a) \wedge (\overline{a} \vee b) \wedge (\overline{a} \vee c)$. After the conventional oracle $O$, the state of $\ket{\vec{v}}: \ket{a} \ket{b} \ket{c}$ becomes
\vspace{-1mm}
{
\small
\[
\frac{1}{\sqrt{8}} (\ket{000} + \ket{001} + \ket{010} + \ket{011}+ \ket{100} + \ket{101} + \ket{110} - \ket{111}),
\]
}
where $\ket{111}$ has a ``$-1$'' phase because it is the solution to formula $\mathcal{F}$. In our approach, the input vector $\ket{\vec{v}}$ is equivalently expanded into $\ket{\vec{v^e}}: \ket{a_{[e_1]}} \ket{a_{[e_2]}} \ket{a_{[e_3]}} \ket{b} \ket{c}$. After applying our parallel oracle $\mathcal{O}$, the state of the input vector $\ket{\vec{v^e}}$ becomes
\vspace{-1mm}
{
\small
\[
\frac{1}{\sqrt{8}} (\ket{\mathbf{\tilde{0}}00} + \ket{\mathbf{\tilde{0}}01} + \ket{\mathbf{\tilde{0}}10} + \ket{\mathbf{\tilde{0}}11} + \ket{\mathbf{\tilde{1}}00} + \ket{\mathbf{\tilde{1}}01} + \ket{\mathbf{\tilde{1}}10} - \ket{\mathbf{\tilde{1}}11}),
\]
} 
where 
$\mathbf{\tilde{0}}$ denotes $000$, $\mathbf{\tilde{1}}$ denotes $111$, 
and $\ket{\mathbf{\tilde{1}}11}$ is the solution to the expanded formula $\mathcal{F}^e: (a_{[e_1]}) \wedge (\overline{a_{[e_2]}} \vee b) \wedge (\overline{a_{[e_3]}} \vee c)$.

\subsection{Parallel Diffuser} \label{subsec:ParallelDiffuser}

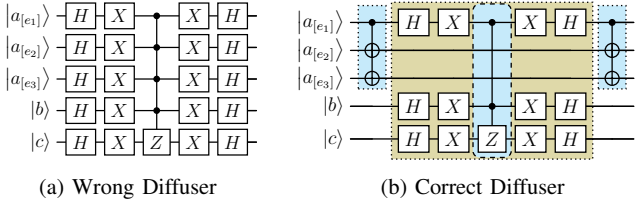
\begin{figure}[tb]
\centering
\begin{subfigure}[b]{0.4\linewidth}
\scalebox{0.6}{
\begin{quantikz}[row sep={7mm,between origins}, column sep=2mm, font=\Large]
\lstick{$\ket{a_{[e_1]}}$} & \gate{H} & \gate{X} & \ctrl{1} & \gate{X} & \gate{H} & \qw \\
\lstick{$\ket{a_{[e_2]}}$} & \gate{H} & \gate{X} & \ctrl{1} & \gate{X} & \gate{H} & \qw \\
\lstick{$\ket{a_{[e_3]}}$} & \gate{H} & \gate{X} & \ctrl{1} & \gate{X} & \gate{H} & \qw \\
\lstick{\ket{b}} & \gate{H} & \gate{X} & \ctrl{1} & \gate{X} & \gate{H} & \qw \\
\lstick{\ket{c}} & \gate{H} & \gate{X} & \gate{Z} & \gate{X} & \gate{H} & \qw
\end{quantikz}
}
\caption{Wrong Diffuser} \label{fig:WrongParallelDiffuser}
\end{subfigure}
~
\begin{subfigure}[b]{0.56\linewidth}
\scalebox{0.62}{
\begin{quantikz}[row sep={7mm,between origins}, column sep=2mm, font=\Large]
\lstick{$\ket{a_{[e_1]}}$} & [1mm]\ctrl{2} \gategroup[3,steps=1,style={dotted,fill=cyan!20, inner xsep=1pt, inner ysep=-1pt}, background,label style={label position=above,anchor=south,yshift=-2mm}]{} & [2mm]\gate{H} \gategroup[5,steps=5,style={dotted,fill=olive!30, inner xsep=1pt, inner ysep=1pt}, background,label style={label position=above,anchor=south,yshift=-2mm}]{} & \gate{X} & \ctrl{3} \gategroup[5,steps=1,style={dashed, rounded corners,fill=cyan!20, inner xsep=0pt, inner ysep=0pt}, background,label style={label position=above,anchor=south,yshift=-0.2cm}]{} & \gate{X} & \gate{H} & [2mm]\ctrl{2} \gategroup[3,steps=1,style={dotted,fill=cyan!20, inner xsep=1pt, inner ysep=-1pt}, background,label style={label position=above,anchor=south,yshift=-2mm}]{} & [1mm]\qw \\[-1mm]
\lstick{$\ket{a_{[e_2]}}$} & \targ{} & \qw & \qw & \qw & \qw & \qw & \targ{} & \qw \\[-1mm]
\lstick{$\ket{a_{[e_3]}}$} & \targ{} & \qw & \qw & \qw & \qw & \qw & \targ{} & \qw \\[-1mm]
\lstick{\ket{b}} & \qw & \gate{H} & \gate{X} & \ctrl{1} & \gate{X} & \gate{H} & \qw & \qw \\
\lstick{\ket{c}} & \qw & \gate{H} & \gate{X} & \gate{Z} & \gate{X} & \gate{H} & \qw & \qw
\end{quantikz}
}
\caption{Correct Diffuser} \label{fig:CorrectParallelDiffuser}
\end{subfigure}
\caption{A parallel diffuser for $\mathcal{F}$.} \label{fig:ParallelDiffuserExample}
\end{figure}

The purpose of the diffuser is to amplify the amplitude of the solution vectors to increase/maximize the probability of the solution being measured. The classic diffuser used in Grover's algorithm adopts the so called {\em inversion about the average} approach to achieve this goal. However, the classic diffuser does not work directly in our parallel setting. Let us use the running example $\mathcal{F}: (a) \wedge (\overline{a} \vee b) \wedge (\overline{a} \vee c)$ again for illustration. Fig.~\ref{fig:WrongParallelDiffuser} shows the case when the classic diffuser is directly applied for $\mathcal{F}$ on all qubits (including the expanded ones $\ket{a_{[e_2]}}$ and $\ket{a_{[e_3]}}$), which generates the wrong result. This is because the classic diffuser assumes all the combinations of the input values have equal probability to occur, i.e., $\ket{a_{[e_1]}} \ket{a_{[e_2]}} \ket{a_{[e_3]}}$ could be $\ket{000}$, $\ket{001}$, $\ket{010}$, $\ldots, \ket{111}$ with equal probability $\frac{1}{8}$. This violates the invariant we want to preserve at all times, i.e., $\ket{a_{[e_1]}} \ket{a_{[e_2]}} \ket{a_{[e_3]}}$ can only be either $\ket{000}$ or $\ket{111}$. The correct parallel diffuser for $\mathcal{F}$ should be the one shown in Fig.~\ref{fig:CorrectParallelDiffuser}. 

Now, let us see what adjustment should be done to utilize the classic diffuser in our parallel setting. Here, we omit the detail of the classic diffuser, which is out of scope of this work. Instead, let us assume that $\ket{\vec{v}}$ in $\mathcal{F}$ is amplified as $(\alpha_0 \ket{000} + \alpha_1 \ket{001} + \alpha_2 \ket{010} + \alpha_3 \ket{011}+ \alpha_4 \ket{100} + \alpha_5 \ket{101} + \alpha_6 \ket{110} + \alpha_7 \ket{111})$, where $\alpha_i \in \mathbb{C}$, $i \in \{1,2,\ldots, 7\}$, and $\sum_{i=0}^7 |\alpha_i|^2 = 1$.

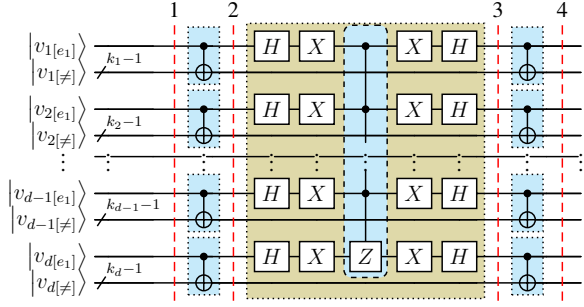
\begin{figure}[tb]
\centering
\scalebox{0.7}{
\begin{quantikz}[row sep={5mm,between origins}, column sep=2mm, font=\Large]
\lstick{$\ket{v_{1[e_1]}}$} & \qw & [6mm]\qw \slice{1} & [6mm]\ctrl{1} \gategroup[2,steps=1,style={dotted,fill=cyan!20, inner xsep=1pt, inner ysep=-1pt}, background,label style={label position=above,anchor=south,yshift=-2mm}]{} & [-2mm]\qw \slice{2} & [6mm]\gate{H} \gategroup[9,steps=5,style={dotted,fill=olive!30, inner xsep=1pt, inner ysep=1pt}, background,label style={label position=above,anchor=south,yshift=-2mm}]{} & \gate{X} & [1mm]\ctrl{2} \gategroup[8,steps=1,style={dashed, rounded corners,fill=cyan!20, inner xsep=0pt, inner ysep=0pt}, background,label style={label position=above,anchor=south,yshift=-0.2cm}]{} & [1mm]\gate{X} & \gate{H} & [-2mm]\qw \slice{3} & [6mm]\ctrl{1} \gategroup[2,steps=1,style={dotted,fill=cyan!20, inner xsep=1pt, inner ysep=-1pt}, background,label style={label position=above,anchor=south,yshift=-2mm}]{} & \qw \slice{4} & [4mm]\qw \\
\lstick{$\ket{v_{1[\neq]}}$} & \qw \qwbundle{k_1-1} & \qw & \targ{} & \qw & \qw & \qw & \qw & \qw & \qw & \qw & \targ{} & \qw & \qw \\[2mm]
 \lstick{$\ket{v_{2[e_1]}}$} & \qw & \qw & \ctrl{1} \gategroup[2,steps=1,style={dotted,fill=cyan!20, inner xsep=1pt, inner ysep=-1pt}, background,label style={label position=above,anchor=south,yshift=-2mm}]{} & \qw & \gate{H} & \gate{X} & \ctrl{2} & \gate{X} & \gate{H} & \qw & \ctrl{1} \gategroup[2,steps=1,style={dotted,fill=cyan!20, inner xsep=1pt, inner ysep=-1pt}, background,label style={label position=above,anchor=south,yshift=-2mm}]{} & \qw & \qw \\
\lstick{$\ket{v_{2[\neq]}}$} & \qw \qwbundle{k_2-1} & \qw & \targ{} & \qw & \qw & \qw & \qw & \qw & \qw & \qw & \targ{} & \qw & \qw \\[-1mm]
 \hspace{-12mm} \vdots & \vdots & & \vdots & & \vdots & \vdots & \vdots & \vdots & \vdots & & \vdots & & \vdots \\[2mm]
  \lstick{$\ket{v_{d-1[e_1]}}$} & \qw & \qw & \ctrl{1} \gategroup[2,steps=1,style={dotted,fill=cyan!20, inner xsep=1pt, inner ysep=-1pt}, background,label style={label position=above,anchor=south,yshift=-2mm}]{} & \qw & \gate{H} & \gate{X} & \ctrl{2} \vqw{-1} & \gate{X} & \gate{H} & \qw & \ctrl{1} \gategroup[2,steps=1,style={dotted,fill=cyan!20, inner xsep=1pt, inner ysep=-1pt}, background,label style={label position=above,anchor=south,yshift=-2mm}]{} & \qw & \qw \\
\lstick{$\ket{v_{d-1[\neq]}}$} & \qw \qwbundle{k_{d-1}-1} & \qw & \targ{} & \qw & \qw & \qw & \qw & \qw & \qw & \qw & \targ{} & \qw & \qw \\[2mm]
  \lstick{$\ket{v_{d[e_1]}}$} & \qw & \qw & \ctrl{1} \gategroup[2,steps=1,style={dotted,fill=cyan!20, inner xsep=1pt, inner ysep=-1pt}, background,label style={label position=above,anchor=south,yshift=-2mm}]{} & \qw & \gate{H} & \gate{X} & \gate{Z} & \gate{X} & \gate{H} & \qw & \ctrl{1} \gategroup[2,steps=1,style={dotted,fill=cyan!20, inner xsep=1pt, inner ysep=-1pt}, background,label style={label position=above,anchor=south,yshift=-2mm}]{} & \qw & \qw \\
\lstick{$\ket{v_{d[\neq]}}$} & \qw \qwbundle{k_{d}-1} & \qw & \targ{} & \qw & \qw & \qw & \qw & \qw & \qw & \qw & \targ{} & \qw & \qw
\end{quantikz}
}
\caption{The parallel diffuser scheme.} \label{fig:ParallelDiffuser}
\end{figure}

Our parallel diffuser is designed to achieve the same effect, i.e., to applify $\ket{\vec{v^e}}$ in formula $\mathcal{F}^e$ as $(\alpha_0 \ket{\mathbf{\tilde{0}}00} + \alpha_1 \ket{\mathbf{\tilde{0}}01} + \alpha_2 \ket{\mathbf{\tilde{0}}10} + \alpha_3 \ket{\mathbf{\tilde{0}}11} + \alpha_4 \ket{\mathbf{\tilde{1}}00} + \alpha_5 \ket{\mathbf{\tilde{1}}01} + \alpha_6 \ket{\mathbf{\tilde{1}}10} + \alpha_7 \ket{\mathbf{\tilde{1}}11})$, where $\mathbf{\tilde{0}}$ denotes $000$, $\mathbf{\tilde{1}}$ denotes $111$. Fig.~\ref{fig:ParallelDiffuser} shows the quantum circuit construction for the proposed parallel diffuser. Suppose a CNF formula $F$ is over $V$, where $|V| = d$. For each variable $v_j \in V$ for $j \in \{1,2, \ldots, d\}$, if $v_j$ appears in $k_j$ clauses in $F$, we use the following notation
\vspace{-1mm}
\[
\ket{v_{j[\neq]}} = \ket{v_{j[e_2]}} \ket{v_{j[e_3]}} \cdots \ket{v_{j[e_{k_j}]}}
\] 
to denote the tensor product of all expanded variables except $v_{j[e_1]}$.
In Step~$1$ of Fig.~\ref{fig:ParallelDiffuser}, each shared variable $\ket{v_j}_1$ is entangled with its expanded variables, i.e., $\ket{v_j}_1 \ket{v_{j[\neq]}}_1 = \alpha_j \ket{0}^{\otimes_{k_j}} + \beta_j \ket{1}^{\otimes_{k_j}}$, where $\alpha_j, \beta_j \in \mathbb{C}$. 

In Step~$2$, each expanded variable is disentangled with $\ket{v_j}$ by a CNOT gate with one control (i.e, $\ket{v_j}$) and $(k_j - 1)$ targets (i.e., $\ket{v_{j[\neq]}}$). Thus, $\ket{v_j}_2 \ket{v_{j[\neq]}}_2 = (\alpha_j \ket{0} + \beta_j \ket{1}) \otimes \ket{0}^{\otimes_{k_j - 1}}$, i.e., $\ket{v_{j[e_q]}}_2$ becomes $\ket{0}$ and is independent from $\ket{v_{j[e_1]}}$ for $q \in \{2, 3, \ldots, k_j\}$.
\tys{This is experimentally possible in circuit QED, in general we can safely say that multi-control multi-target CNOT type gate can be realized. However whether there is any constraints under the distributive setting is not clear to me at this point.}

In Step~$3$, only $\ket{v_{j[e_1]}}_2$ is selected as the representative for the diffusion process for all $j \in \{1,2, \ldots, d\}$, and the classic diffuser can be utilized. Actually, the selected representatives $\ket{v_{1[e_1]}}_2 \ket{v_{2[e_1]}}_2 \cdots \ket{v_{d[e_1]}}_2$ are exactly the input of the classic diffuser $\ket{v_1} \ket{v_2} \cdots \ket{v_d}$, as shown in Fig.~\ref{fig:ClassicDiffuser}.

Assume $\ket{v_{j[e_1]}}_3$ is amplified as $\alpha_j' \ket{0} + \beta_j' \ket{1}$ after the diffusion process. In Step~$4$, the expanded variables $\ket{v_{j[\neq]}}$ are entangled back with $\ket{v_j}$ by a CNOT gate with one control (i.e, $\ket{v_j}$) and $(k_j - 1)$ targets (i.e., $\ket{v_{j[\neq]}}$). Thus, $\ket{v_j}_4 \ket{v_{j[\neq]}}_4 = \alpha_j' \ket{0}^{\otimes_{k_j}} + \beta_j' \ket{1}^{\otimes_{k_j}}$. Theorem~\ref{thm:ParallelDiffuser} shows the details step by step and proves that our parallel diffuser has the same effect as the classic diffuser.




\begin{theorem}[Parallel Diffuser Correctness] \label{thm:ParallelDiffuser}
Let $\ket{\vec{v}}: \ket{v_1} \ket{v_2} \cdots \ket{v_d}$ be the input vector of $F$ and $D$ be the classic diffuser such that
\[
D(\ket{\vec{v}}) = \sum_{i=0}^{2^d-1} \alpha_i' \left(\ket{b_1} \ket{b_2} \cdots \ket{b_d} \right), 
\]
where the index $i$ is represented as the binary string $\ket{b_1} \ket{b_2} \cdots \ket{b_d} \in \{0,1\}^d$.
If the input vector of $F^e$ is $\ket{\vec{v^e}}$, our parallel diffuser $\mathcal{D}$ ensures the following:
\[
\mathcal{D}(\ket{\vec{v^e}}) = \sum_{i=0}^{2^d-1} \alpha_i' \left( \ket{b_1}^{\otimes_{k_1}} \ket{b_2}^{\otimes_{k_2}} \cdots \ket{b_d}^{\otimes_{k_d}} \right).
\]
\end{theorem}

\begin{proof}
We prove this theorem based on the steps shown in Fig.~\ref{fig:ParallelDiffuser}. Let $\ket{\vec{v}} = \sum_{i = 0}^{2^d-1} \alpha_i (\ket{b_1} \ket{b_2} \cdots \ket{b_d})$ be the input of the classic diffuser $D$, and $\ket{\vec{v^e}} = \sum_{i=0}^{2^d-1} \alpha_i (\ket{b_1}^{\otimes_{k_1}} \ket{b_2}^{\otimes_{k_2}} \cdots \ket{b_d}^{\otimes_{k_d}})$ be the input of our parallel diffuser $\mathcal{D}$.

Initially, $\ket{\vec{v^e}}_1 = \sum_{i=0}^{2^d-1} \alpha_i (\ket{b_1}^{\otimes^{k_1}} \ket{b_2}^{\otimes^{k_2}} \cdots \ket{b_d}^{\otimes^{k_d}})$. In step~$2$, $\ket{v_{j[\neq]}}$ is disentangled from $\ket{v_{j[e_1]}}$ and becomes $\ket{0}^{\otimes_{k_j - 1}}$ for each $j \in \{1,2, \ldots, d\}$. Thus, we have
\vspace{-1mm}
\[
\ket{\vec{v^e}}_2 = \sum_{i=0}^{2^d-1} \alpha_i \left( \ket{b_1}\ket{0}^{\otimes_{k_1 - 1}} \ket{b_2}\ket{0}^{\otimes_{k_2 - 1}} \cdots \ket{b_d} \ket{0}^{\otimes_{k_d - 1}} \right).
\]
Since $\ket{v_{1[\neq]}}_2$ is now $\ket{0}^{\otimes_{k_1 - 1}}$ and is independent of other terms, we can reorder the sequence of qubits in $\ket{\vec{v^e}}_2$ to move $\ket{v_{1[\neq]}}$ to the end so that we can
move it out from the summation and have
\vspace{-1mm}
\[
\ket{\vec{v^e}}_2 \!=\!\! \left( \sum_{i=0}^{2^d-1} \alpha_i (\ket{b_1} \ket{b_2}\ket{0}^{\otimes_{k_2 - 1}} \cdots \ket{b_d} \ket{0}^{\otimes_{k_d - 1}}) \right) \otimes \ket{0}^{\otimes_{k_1 - 1}}. 
\]
We can do the reordering and rewriting recursively for $\ket{v_{j[\neq]}}$ starting from $j=1$ to $d$. Then, we have
\vspace{-1mm}
{
\small
\begin{eqnarray}
\ket{\vec{v^e}}_2 \!\!\!\!\!\!& = &\!\!\!\!\!\! \left( \sum_{i=0}^{2^d-1} \alpha_i (\ket{b_1} \ket{b_2} \cdots \ket{b_d}) \right) \!\!\otimes \ket{0}^{\otimes_{k_1 - 1}} \ket{0}^{\otimes_{k_2 - 1}} \cdots \ket{0}^{\otimes_{k_d - 1}} \nonumber \\
\!\!\!\!& = &\!\!\!\! \ket{\vec{v}} \otimes \ket{0}^{\otimes_{k_1 - 1}} \ket{0}^{\otimes_{k_2 - 1}} \cdots \ket{0}^{\otimes_{k_d - 1}} \nonumber. 
\end{eqnarray}
}
In Step~$3$, the classic diffuser $D$ is applied on $\ket{\vec{v}}$. Thus, we have
\vspace{-1mm}
{
\small
\[
\ket{\vec{v^e}}_3 \!=\! \left( \sum_{i=0}^{2^d-1} \alpha_i' (\ket{b_1} \ket{b_2} \cdots \ket{b_d}) \right) \otimes \ket{0}^{\otimes_{k_1 - 1}} \ket{0}^{\otimes_{k_2 - 1}} \cdots \ket{0}^{\otimes_{k_d - 1}} \nonumber
\]
}
Now, let us reorder the sequence of qubits again to move $\ket{v_{j[\neq]}}$ right after $\ket{v_j}$ for all 
$j \in \{ 1,2, \ldots, d \}$ such that we can bring $\ket{0}^{\otimes_{k_j-1}}$ back in the summation and have
\vspace{-1mm}
\[
\ket{\vec{v^e}}_3 = \sum_{i=0}^{2^d-1} \alpha_i' \left( \ket{b_1}\ket{0}^{\otimes_{k_1 - 1}} \ket{b_2}\ket{0}^{\otimes_{k_2 - 1}} \cdots \ket{b_d}\ket{0}^{\otimes_{k_d - 1}} \right)
\]
In Step~$4$, $\ket{v_{j[\neq]}}$ is entangled back (in the same state) with $\ket{v_{j[e_1]}}$ for all $j \!\in\! \{1,2, \ldots, d\}$. We have the following to finish the proof.
\vspace{-1mm}
\[
\ket{\vec{v^e}}_4 = \mathcal{D}(\ket{\vec{v^e}}) = \sum_{i=0}^{2^d-1} \alpha_i' \left( \ket{b_1}^{\otimes_{k_1}} \ket{b_2}^{\otimes_{k_2}} \cdots \ket{b_d}^{\otimes_{k_d}} \right) \vspace{-2mm}
\]
\end{proof}

\subsection{Analysis and Simulation} \label{subsec:ParallelSATEvaluation}


Now, we theoretically compare the time complexity of our parallel quantum SAT solver with the conventional (sequential) quantum SAT solver. A Grover iteration includes one oracle process and one diffuser process. Given a formula $F$ with $m$ clauses, the conventional oracle takes $O(m)$ time complexity to mark the solutions with ``$-1$'' phases, while our parallel oracle only takes $O(1)$ constant time to do so. For the diffuser, both conventional and parallel versions take $O(1)$ as they do not depend on the number of clauses. Thus, for one Grover iteration, the conventional version takes $O(m)$ linear time, while our parallel version only takes $O(1)$ constant time. 

What about the number of iterations required for our parallel Grover iteration to obtain the solutions? The answer is $O(\sqrt{N/M})$, the same as that of the conventional one, where $N$ is the size of the search space and $M$ is the number of solutions (c.f.~Section~\ref{subsec:Grover}). Notice that although additional expanded variables are introduced, they have the same values and are entangled with the original variables. In addition, only original variables are involved in the diffusion process. Thus, the size of the search space remains the same. Overall, the conventional (sequential) quantum SAT solver takes $O(m) \cdot O(\sqrt{N/M})$ time complexity, while our parallel approach only takes $O(\sqrt{N/M})$, which brings an linear time improvement.

\begin{figure}[tb]
\centering
\scalebox{0.85}{
\begin{tikzpicture}
\begin{axis}[  
    ybar,  
    enlargelimits=0.1,  
    ylabel={Counts}, 
    xlabel={Measured Outputs ($|abc\rangle$)},  
    symbolic x coords={000, 001, 010, 011, 100, 101, 110, 111}, 
    xtick=data,  
    nodes near coords, 
    nodes near coords align={vertical},
    ymin=0,
    enlarge y limits=upper,
    height=3.2cm,
    width=9cm,
    bar width=12pt,
]  
\addplot coordinates {(000,269) (001,245) (010,256) (011,258) (100,252) (101,283) (110,240) (111,6389)};  
\end{axis}
\end{tikzpicture}
}
\caption{Simulation result of our parallel approach for formula $\mathcal{F}$.}
\label{fig:Parallel_SAT_Result}
\end{figure}
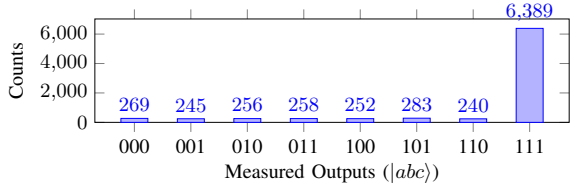

We have implemented the parallel SAT solver for our running example $\mathcal{F}: \!(a) \wedge (\overline{a} \vee b) \wedge (\overline{a} \vee c)$ in Qiskit~\cite{GT19}. The implementation can be obtained in~\cite{Parallel_SAT}. Totally, nine qubits are required (three for variable $a$, two for variables $b$ and $c$, three for all the clauses, and one for formula $\mathcal{F}$). Only one Grover iteration is required. Fig.~\ref{fig:Parallel_SAT_Result} shows the simulation result of performing Grover's algorithm for $8,192$ shots. The x-axis shows the measured outcome of $\ket{abc}$, while the y-axis shows the count of each outcome being measured. One can observe that $\ket{111}$, the solution to formula $\mathcal{F}$, has overwhelming higher probability over other non-solution inputs that are almost negligible. Experimentally, this also confirms the correctness of our parallel quantum SAT solving technique.


\section{Distributed Quantum SAT Solver} \label{sec:DistributedSAT}

In this section, we consider the scenario where one quantum computer has insufficient qubits to handle the whole SAT problem. To overcome this issue, we follow the ``divide and conquer'' strategy and develop a {\em distributed} quantum SAT solver, including a distributed oracle (Section~\ref{subsec:DistributedOracle}) and a distributed diffuser (Section~\ref{subsec:DistributedDiffuser}).

\subsection{Distributed Oracle} \label{subsec:DistributedOracle}

Let us recall the design of our parallel oracle in Fig.~\ref{fig:ParallelOracle}. The quantum circuit for processing each clause $C_i^e$ is independent of each other for $i \in \{1,2,\ldots, m\}$ and thus can be naturally handled by one dedicated quantum computer. The critical question here is ``how to handle the conjunction distributedly'', i.e., how to distributedly perform the CNOT gate with $m$ control qubits and one target qubit.

\begin{figure}[tb]
\centering
\scalebox{0.65}{
\tikzset{
my label/.append style={above right,xshift=1mm}
}
\begin{quantikz}[row sep={6mm,between origins}, column sep=5mm, font=\Large]
\lstick{\ket{C_1}} & [-3mm] \ctrl{1} & [-3mm] \qw & [-5mm] \qw & [-3mm] \qw \slice{1} & \qw \slice{2} & [2mm] \gate{Z} & [-6mm] \qw & [-5mm] \qw \slice{3} & \qw \\
\lstick{\ket{e_1}} & \targ{} & \meter{0/1} \vcw{7} & \qw & \qw & \qw & \qw & \qw & \qw & \qw \\[-1mm]
\wave&&&&&&&&&\\[-1mm]
\lstick{\vdots} & & & & \vdots & & & & \vdots & \\[-1mm]
 \wave&&&&&&&&&\\
\lstick{\ket{C_m}} & \qw & \qw & \ctrl{1} & \qw & \qw & \qw & \qw & \gate{Z} & \qw \\
\lstick{\ket{e_m}} & \qw & \qw & \targ{} & \meter{0/1} \vcw{4} & \qw & \qw & \qw & \qw & \qw \\
\wave[fill=cyan!10]&&&&&&&&& \\[1mm]
\lstick{\ket{\widehat{e_1}}} & \qw & \gate{X} & \qw & \qw & \ctrl{1} \gategroup[4,steps=1,style={dashed,
rounded corners,fill=cyan!20, inner xsep=0.1pt}, background,label style={label position=below,anchor=north,yshift=-2mm}]{{}} & \meter[style={label position=below right}]{\ket{\pm}} \vcw{-8} & \qw & \qw & \qw \\
\hspace{-8mm} \vdots & & & \cdots  & & \vdots & & \cdots & & \\
\lstick{\ket{\widehat{e_m}}} & \qw & \qw & \qw & \gate{X} & \ctrl{1} \vqw{-1} & \qw & \qw & \meter[style={label position=below right}]{\ket{\pm}} \vcw{-5} & \qw \\
\lstick{\ket{t}} & \qw & \qw & \qw & \qw & \gate{U} & \qw & \qw & \qw & \rstick{\ket{t'}} \qw
\end{quantikz}
\hspace{-2mm}
=
\begin{quantikz}[row sep={7mm,between origins}, column sep=2mm, font=\Large]
\lstick{\ket{C_1}} & \ctrl{1} \gategroup[4,steps=1,style={dashed,
rounded corners,fill=cyan!20, inner xsep=0.1pt}, background,label style={label position=below,anchor=north,yshift=-2mm}]{{}} & \qw \\
\hspace{-9mm} \vdots & \vdots \\
\lstick{\ket{C_m}} & \ctrl{1} \vqw{-1} & \qw \\
\lstick{\ket{t}} & \gate{U} & \qw
\end{quantikz}
}
\caption{The distributed $m$-controlled-$U$ gate scheme.}\label{fig:mCUgate}
\end{figure}
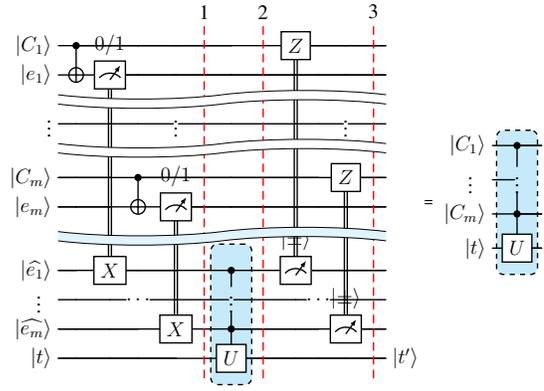

Sarvaghad‐Moghaddam and Zomorodi proposed a general protocol for distributed quantum gates~\cite{SZ21} based on quantum teleportation~\cite{BBC93,BPM97, NKL98,RHR04}. However, the correctness of the protocol was not proved in their paper. Inspired by their work, we develop a protocol for the distributed controlled-$U$ gate,
where $U$ is an arbitrary quantum gate, and further prove its correctness. Fig.~\ref{fig:mCUgate} shows the design of the protocol. Suppose we want to perform a controlled $U$ gate with $m$ control qubits, as shown in the right side of Fig.~\ref{fig:mCUgate}, where $\ket{C_i}$ is the control qubit for $i \in \{1,2,\ldots, m\}$ and $\ket{t}$ is the target qubit. The proposed distributed protocol is designed in a way that the $m$ control qubits need not be in the same quantum computer (node) where the target qubit $\ket{t}$ is located. Let us assume that the control qubit $\ket{C_i}$ is located on node $i$ where $i \in \{1,2,\ldots, m\}$, and the target qubit $\ket{t}$ is located on a master node, as shown in the left side of Fig.~\ref{fig:mCUgate}. 
To perform the controlled $U$ gate remotely, initially, each node $i$ shares, with the master node, a pair of the following entangled qubits:
\vspace{-1mm}
\[
\ket{e_i} \ket{\widehat{e_i}} = \frac{1}{\sqrt{2}} (\ket{00} + \ket{11}) \mbox{, for all } i \in \{1,2,\ldots, m\},
\]
where node $i$ holds qubit $\ket{e_i}$ and node master holds qubit $\ket{\widehat{e_i}}$.
\tys{the notation above is incorrect as \ket{e_i} \ket{\widehat{e_i}} are separable not entangled, should be written as \ket{e_i\widehat{e_i}} .}

In step~$1$, each node $i$ performs a CNOT gate on $\ket{C_i} \ket{e_i}$, measures qubit $\ket{e_i}$ in the standard ($\ket{0}$ and $\ket{1}$) basis and then sends the measurement outcome to node master via a non-quantum channel 
(e.g., TCP/IP, etc.).
After receiving the measurement outcome, the node master applies an $X$ gate on qubit $\ket{\widehat{e_i}}$ if the measurement outcome is $\ket{1}$; otherwise, nothing is performed. After this step, the qubit $\ket{e_i}$ collapses, and the two qubits $\ket{C_i} \ket{\widehat{e_i}}$ become entangled in the same state, i.e., they are either in state $\ket{00}$ or $\ket{11}$.

In step~$2$, since $\ket{C_i}$ and $\ket{\widehat{e_i}}$ have the same state, applying the controlled $U$ gate with $\ket{\widehat{e_i}}$ as the $m$ control qubits is equivalent to that with $\ket{C_i}$ as the $m$ control qubits for $i \in \{1,2,\ldots, m\}$.

Step~$3$ disentangles $\ket{C_i}$ from $\ket{\widehat{e_i}}$. To do so, the node master measures the qubit $\ket{\widehat{e_i}}$ in the $\ket{+}$ and $\ket{-}$ basis and then sends the measurement outcome to node $i$ via a non-quantum channel. After receiving the measurement outcome, node $i$ performs a $Z$ gate on qubit $\ket{C_i}$ if the outcome is $\ket{-}$; otherwise, nothing is performed. Once node $i$ finishes this step for each $i \in \{1,2,\ldots, m\}$, the operation of the controlled $U$ gate is accomplished distributedly among the $m+1$ nodes. Theorem~\ref{thm:ProtocalCorrectness} shows the details step by step and proves the correctness of the distributed protocol.

\begin{theorem}[Distributed Protocol Correctness of Fig.~\ref{fig:mCUgate}] \label{thm:ProtocalCorrectness} 
\mbox{} \\
Let $\ket{C_i} = x_i \ket{0} + y_i \ket{1}$, where $x_i, y_i \in \mathbb{C}$ and $i \in \{1,,2, \ldots, m\}$. (1) In Step~$1$, $\ket{\widehat{e_i}}\ket{C_i} = x_i \ket{00} + y_i \ket{11}$ for all $i \in \{1,2, \ldots, m \}$. \\
(2) In Step~$2$, $\ket{t'} = U(\ket{t})$ iff $\ket{C_i} = \ket{1}$ for all $i \in \{1,2, \ldots, m\}$. \\
(3) In Step~$3$, $\ket{C_i} = x_i \ket{0} + y_i \ket{1}$ for all $i \in  \{1,2, \ldots, m\}$.
\end{theorem}

\begin{proof}
We prove each step of the distributed protocol as follows.

(1). Initially, $\ket{e_i \widehat{e_i}}_0 \ket{C_i}_0 = \frac{1}{\sqrt{2}}(\ket{00} + \ket{11}) \otimes (x_i \ket{0} + y_i \ket{1})$ $= \frac{1}{\sqrt{2}}(x_i \ket{000} + y_i \ket{001} + x_i \ket{110} + y_i \ket{111})$. After the CNOT gate, $\mbox{CNOT}(\ket{e_i \widehat{e_i}}_0 \ket{C_i}_0) = \frac{1}{\sqrt{2}}(x_i \ket{000} + y_i \ket{101} + x_i \ket{110} + y_i \ket{011})$
$= \frac{\ket{0}}{\sqrt{2}} \bigl( x_i \ket{00} + y_i \ket{11} \bigr) + \frac{\ket{1}}{\sqrt{2}} \bigl( x_i \ket{10} + y_i \ket{01} \bigr)$. If we measure $\ket{e_i}$ now, it has $\frac{1}{2}$ probability to be $\ket{0}$ and $\frac{1}{2}$ probability to be $\ket{1}$. We examine these two cases below.

Case~$(a)$: the measurement outcome is $\ket{0}$. In this case, $\ket{\widehat{e_i}}_1 \ket{C_i}_1$ collapses to the state of $(x_i \ket{00} + y_i \ket{11})$, and we do not perform any operation on $\ket{\widehat{e_i}}$. Thus, $\ket{e_i}_1 \ket{C_i}_1 = x_i \ket{00} + y_i \ket{11}$.

Case~$(b)$: the measurement outcome is $\ket{1}$. In this case, $\ket{\widehat{e_i}}_1 \ket{C_i}_1$ collapses to the state of $(x_i \ket{10} + y_i \ket{01})$, and we apply an $X$ gate on $\ket{\widehat{e_i}}$. Thus, $\ket{\widehat{e_i}}_1 \ket{C_i}_1 = x_i \ket{00} + y_i \ket{11}$.


(2). In Step~$2$, $\ket{t}_2 = U(\ket{t}_0)$ iff $\ket{\widehat{e_i}}_1 = \ket{1}$ for all $i \in \{1,2, \ldots, m\}$. Based on~$(1)$ we just proved, $\ket{C_i}_1$ is entangled with $\ket{\widehat{e_i}}_1$ in the same state, i.e., $\ket{C_i}_1 = \ket{\widehat{e_i}}_1$. Thus, we can conclude that $\ket{t'} = U(\ket{t})$ iff $\ket{C_i} = \ket{1}$ for all $i \in \{1,2, \ldots, m\}$.

(3). After Step~$1$, $\ket{\widehat{e_i}} \ket{C_i} = x_i \ket{00} + y_i \ket{11}$. Since we are going to measure $\ket{\widehat{e_i}}$ in the $Z$ basis, i.e., $\ket{\pm}$, let us rewrite the state of $\ket{\widehat{e_i}}$ in the $Z$ basis. $\ket{\widehat{e_i}}_2 \ket{C_i}_2 = \bigl( \frac{\ket{+} + \ket{-}}{\sqrt{2}} \bigr) \bigl( x_i \ket{0} \bigr) + \bigl( \frac{\ket{+} - \ket{-}}{\sqrt{2}} \bigr) \bigl(  y_i \ket{1} \bigr)$
$= \frac{\ket{+}}{\sqrt{2}} \bigl( x_i \ket{0} + y_i \ket{1} \bigr) + \frac{\ket{-}}{\sqrt{2}} \bigl( x_i \ket{0} - y_i \ket{1} \bigr)$. If we measure $\ket{\widehat{e_i}}$ in the $Z$ basis now, it has $\frac{1}{2}$ probability to be $\ket{+}$ and $\frac{1}{2}$ probability to be $\ket{-}$. Thus, there are two cases.

Case~$(a)$: the measurement outcome is $\ket{+}$. In this case, $\ket{C_i}_3$ collapses to the state of $(x_i \ket{0} + y_i \ket{1})$, and nothing is performed on $\ket{C_i}_3$. Thus, $\ket{C_i}_3 = x_i \ket{0} + y_i \ket{1}$.

Case~$(b)$: the measurement outcome is $\ket{-}$. In this case, $\ket{C_i}_3$ collapses to the state of $(x_i \ket{0} - y_i \ket{1})$, and then a $Z$ gate is applied on $\ket{C_i}_3$. Thus, $\ket{C_i}_3 = x_i \ket{0} + y_i \ket{1}$ because $Z(\ket{0}) = \ket{0}$ and $Z(\ket{1}) = - \ket{1}$.
\end{proof}

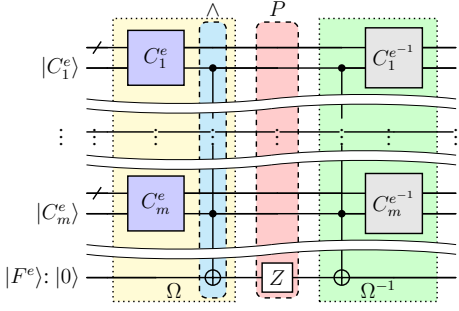
\begin{figure}[tb]
\centering
\scalebox{0.68}{
\begin{quantikz}[row sep={4mm,between origins}, column sep=4mm, font=\Large]
 & \qw \qwbundle{} & \gate[wires=2, style={fill=blue!20}][11mm]{C_1^e} \gategroup[9,steps=2,style={dotted,fill=yellow!20, inner xsep=5pt, inner ysep=2pt}, background,label style={label position=below,anchor=south,yshift=-0.23cm}]{{\sc $\Omega$}} & \qw \gategroup[9,steps=1,style={dashed, rounded corners,fill=cyan!20, inner xsep=0pt, inner ysep=0pt}, background,label style={label position=above,anchor=south,yshift=-0.2cm}]{{\sc $\wedge$}} & [4mm]\qw \gategroup[9,steps=1,style={dashed, rounded corners,fill=red!20, inner xsep=0pt, inner ysep=0pt}, background,label style={label position=above,anchor=south,yshift=-0.2cm}]{{\sc $P$}} & [4mm] \qw \gategroup[9,steps=2,style={dotted,fill=green!20, inner xsep=5pt, inner ysep=2pt}, background,label style={label position=below,anchor=south,yshift=-0.23cm}]{{\sc $\Omega^{-1}$}} & [-1mm]\gate[wires=2, style={fill=gray!20}]{C_1^{e^{-1}}} & [2mm]\qw \\
 \lstick{\ket{C_1^e}} & \qw & \qw & \ctrl{2} & \qw & \ctrl{2} & \qw & \qw \\[3mm]
 \wave &&&&&&&& \\[1.5mm]
  \hspace{-10mm} \vdots & \hspace{-5mm} \vdots & \vdots & \vdots & \vdots & \vdots & \hspace{-2mm} \vdots & \vdots \\[1.5mm]
\wave &&&&&&&& \\[2.5mm]
 & \qw \qwbundle{} & \gate[wires=2, style={fill=blue!20}][11mm]{C_m^e} & \qw & \qw & \qw &[-1mm]\gate[wires=2, style={fill=gray!20}]{C_m^{e^{-1}}}  & \qw \\
 \lstick{\ket{C_m^e}} & \qw & \qw & \ctrl{2} \vqw{-3} & \qw & \ctrl{2} \vqw{-3} & \qw & \qw \\[3mm]
\wave &&&&&&&& \\[1.5mm]
 \lstick{\ket{F^e}: \ket{0}} & \qw & \qw & \targ{} & \gate{Z} & \targ{} & \qw & \qw
\end{quantikz}
}
\caption{The parallel and distributed oracle construction scheme.}\label{fig:ParallelDistributedOracle}
\end{figure}

With this developed protocol, we can perform the conjunction of $m$ clauses distributedly. The design of the distributed oracle is shown in Fig.~\ref{fig:ParallelDistributedOracle}, where each clause $C_i^e$ is handled by node $i$, and the node master interacts with node $i$ on qubit $\ket{C_i^e}$ for all $i \in \{1,2,\ldots, m\}$ as the $m$ control qubits to accomplish the conjunction based on the distributed protocol. Notice that there are two conjunction operations to be performed distributedly: one is in the $\Omega$ block and the other is in the $\Omega^{-1}$ block.

\begin{theorem}
Our distributed oracle is correct.
\end{theorem}
\begin{proof}
The correctness follows from Theorem~\ref{thm:ParallelOracleCorrectness} and Theorem~\ref{thm:ProtocalCorrectness}; the detailed proof is omitted here.
\end{proof}

\begin{figure}[tb]
\centering
\scalebox{0.56}{
\tikzset{
my label/.append style={above right,xshift=1mm}
}
\begin{quantikz}[row sep={5.5mm,between origins}, column sep=1.5mm, font=\Large]
\lstick{$\ket{a_{[e_1]}}$} & [4mm]\gate{X} \gategroup[18,steps=13,style={dotted,fill=yellow!20, inner xsep=7pt, inner ysep=2pt}, background,label style={label position=below,anchor=south,yshift=-0.23cm}]{{\sc $\Omega$}} \gategroup[2,steps=2,style={dashed, rounded corners,fill=blue!20, inner xsep=0pt, inner ysep=0pt}, background,label style={label position=above,anchor=south,yshift=-0.2cm}]{{\sc $C_1^e$}} & \ctrl{1} & [5mm] \qw \gategroup[18,steps=10,style={dashed, rounded corners,fill=cyan!20, inner xsep=5pt, inner ysep=0pt}, background,label style={label position=above,anchor=south,yshift=-0.2cm}]{{\sc $\wedge$}} & \qw & [-4mm]\qw & \qw & [-4mm]\qw & \qw & [-2mm]\qw & \qw & [-3mm]\qw & [-3mm]\qw & [2mm] \qw \gategroup[18,steps=1,style={dashed, rounded corners,fill=red!20, inner xsep=0pt, inner ysep=0pt}, background,label style={label position=above,anchor=south,yshift=-0.2cm}]{{\sc $P$}} & [5mm]\gate[wires=18,style={fill=green!15}][14mm]{\Omega^{-1}} & [2mm]\qw \\[1mm]
\lstick{\ket{C_1^e}:\ket{0}} & \gate{X} & \targ{} & \ctrl{1} & \qw & \qw & \qw & \qw & \qw & \qw & \qw & \qw & \qw & \qw &  & \qw \\[1mm]
\lstick{\ket{e_1}} & \qw & \qw & \targ{} & \meter{0/1} \vcw{12} & \qw & \qw & \qw & \qw & \qw & \gate{Z} & \qw & \qw & \qw & & \qw \\[1mm]
\wave &&&&&&&&&&&&&&& \\[3mm]
\lstick{$\ket{a_{[e_2]}}$} & \qw \gategroup[3,steps=2,style={dashed, rounded corners,fill=blue!20, inner xsep=0pt, inner ysep=0pt}, background,label style={label position=above,anchor=south,yshift=-2.5mm}]{{\sc $C_2^e$}} & \ctrl{1} & \qw & \qw & \qw & \qw & \qw & \qw & \qw & \qw & \qw & \qw & \qw & & \qw \\
\lstick{\ket{b}} & \gate{X} & \ctrl{1} & \qw & \qw & \qw & \qw & \qw & \qw & \qw & \qw & \qw & \qw & \qw &  & \qw \\[1mm]
\lstick{\ket{C_2^e}: \ket{0}} & \gate{X} & \targ{} & \qw & \qw & \ctrl{1} & \qw & \qw & \qw & \qw & \qw & \qw & \qw & \qw & & \qw \\[1mm]
\lstick{\ket{e_2}} & \qw & \qw & \qw & \qw & \targ{} & \meter{0/1} \vcw{8} & \qw & \qw & \qw & \qw & \gate{Z} & \qw & \qw & & \qw \\[1mm]
\wave &&&&&&&&&&&&&&& \\[3mm]
\lstick{$\ket{a_{[e_3]}}$} & \qw \gategroup[3,steps=2,style={dashed, rounded corners,fill=blue!20, inner xsep=0pt, inner ysep=0pt}, background,label style={label position=above,anchor=south,yshift=-2.5mm}]{{\sc $C_3^e$}} & \ctrl{1} & \qw & \qw & \qw & \qw & \qw & \qw & \qw & \qw & \qw & \qw & \qw &  & \qw \\
\lstick{\ket{c}} & \gate{X} & \ctrl{1} & \qw & \qw & \qw & \qw & \qw & \qw & \qw & \qw & \qw & \qw & \qw & & \qw \\[1mm]
\lstick{\ket{C_3^e}: \ket{0}} & \gate{X} & \targ{} & \qw  & \qw & \qw & \qw & \ctrl{1} & \qw & \qw & \qw & \qw & \qw & \qw & & \qw \\[1mm]
\lstick{\ket{e_3}} & \qw & \qw & \qw & \qw & \qw & \qw & \targ{} & \meter{0/1} \vcw{4} & \qw & \qw & \qw & \gate{Z} & \qw &  & \qw \\[1mm]
\wave &&&&&&&&&&&&&& & \\[2mm]
\lstick{\ket{\widehat{e_1}}} & \qw & \qw & \qw & \gate{X} & \qw & \qw & \qw & \qw & \ctrl{1} & \meter[style={label position=below right}]{\ket{\pm}} \vcw{-12} & \qw & \qw & \qw &  & \qw \\[2mm]
\lstick{\ket{\widehat{e_2}}} & \qw & \qw & \qw & \qw & \qw & \gate{X} & \qw & \qw & \ctrl{1} & \qw & \meter[style={label position=below right}]{\ket{\pm}} \vcw{-8} & \qw & \qw &  & \qw \\[2mm]
\lstick{\ket{\widehat{e_3}}} & \qw & \qw & \qw & \qw & \qw & \qw & \qw & \gate{X} & \ctrl{1} & \qw & \qw & \meter[style={label position=below right}]{\ket{\pm}} \vcw{-4} & \qw & & \qw \\
\lstick{\ket{F^e}: \ket{0}} & \qw & \qw & \qw & \qw & \qw & \qw & \qw & \qw & \targ{} & \qw & \qw & \qw & \gate{Z} &  & \qw
\end{quantikz}
}
\caption{A parallel and distributed oracle for $(a) \wedge (\overline{a} \vee b) \wedge (\overline{a} \vee c)$.}\label{fig:ParallelDistributedOracleExample}
\end{figure}
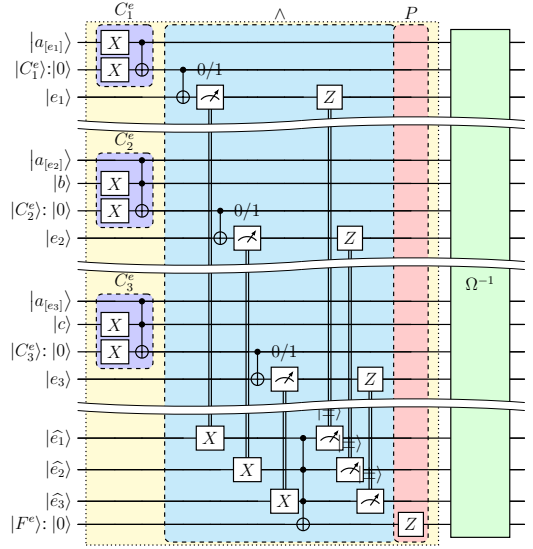

Let us use the running example for illustration. Fig.~\ref{fig:ParallelDistributedOracleExample} shows the distributed oracle for the formula $\mathcal{F}: (a) \wedge (\overline{a} \vee b) \wedge (\overline{a} \vee c)$. Since there are three clauses, we need four nodes involved (one for each clause and one for the master node). Each node $i$ shares the pair $\ket{e_i} \ket{\widehat{e_i}}$ with the node master for $i \in \{1,2,3 \}$ such that node $i$ holds qubit $\ket{e_i}$, while the node master holds qubit $\ket{\widehat{e_i}}$. The conjunction is performed based on the proposed distributed protocol, as shown in the $\wedge$-block in cyan color. The other conjunction operation in the $\Omega^{-1}$ is identical, which is omitted here due to the space limit.
\tys{In this distributive version of oracle, each node must maintain sharing with the master node a pair of entangled qubits at any given time, i,e, either $\ket{e_i\widehat{e_i}} $ or $\ket{c_i\widehat{e_i}}$, at least until the end of the entire Grover iterations. This requirement can potentially become a bottleneck for implementation.}

\subsection{Distributed Diffuser} \label{subsec:DistributedDiffuser}

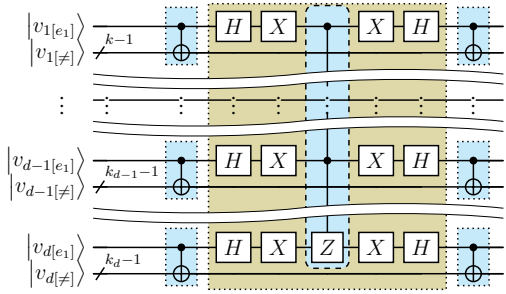
\begin{figure}[tb]
\centering
\scalebox{0.7}{
\begin{quantikz}[row sep={5mm,between origins}, column sep=2mm, font=\Large]
\lstick{$\ket{v_{1[e_1]}}$} & \qw & [10mm]\ctrl{1} \gategroup[2,steps=1,style={dotted,fill=cyan!20, inner xsep=1pt, inner ysep=-1pt}, background,label style={label position=above,anchor=south,yshift=-2mm}]{} & [3mm]\gate{H} \gategroup[10,steps=5,style={dotted,fill=olive!30, inner xsep=1pt, inner ysep=1pt}, background,label style={label position=above,anchor=south,yshift=-2mm}]{} & \gate{X} & [1mm]\ctrl{3} \gategroup[9,steps=1,style={dashed, rounded corners,fill=cyan!20, inner xsep=0pt, inner ysep=0pt}, background,label style={label position=above,anchor=south,yshift=-0.2cm}]{} & [1mm]\gate{X} & \gate{H} & [3mm]\ctrl{1} \gategroup[2,steps=1,style={dotted,fill=cyan!20, inner xsep=1pt, inner ysep=-1pt}, background,label style={label position=above,anchor=south,yshift=-2mm}]{} & [2mm]\qw \\
\lstick{$\ket{v_{1[\neq]}}$} & \qw \qwbundle{k-1} & \targ{} & \qw & \qw & \qw & \qw & \qw & \targ{} & \qw \\
 \wave &&&&&&&&& \\[-1mm]
 \hspace{-12mm} \vdots & \vdots & \vdots & \vdots & \vdots & \vdots & \vdots & \vdots & \vdots & \\
 \wave &&&&&&&&& \\[1.5mm]
  \lstick{$\ket{v_{d-1[e_1]}}$} & \qw & \ctrl{1} \gategroup[2,steps=1,style={dotted,fill=cyan!20, inner xsep=1pt, inner ysep=-1pt}, background,label style={label position=above,anchor=south,yshift=-2mm}]{} & \gate{H} & \gate{X} & \ctrl{3} \vqw{-2} & \gate{X} & \gate{H} & \ctrl{1} \gategroup[2,steps=1,style={dotted,fill=cyan!20, inner xsep=1pt, inner ysep=-1pt}, background,label style={label position=above,anchor=south,yshift=-2mm}]{} & \qw \\
\lstick{$\ket{v_{d-1[\neq]}}$} & \qw \qwbundle{k_{d-1}-1} & \targ{} & \qw & \qw & \qw & \qw & \qw & \targ{} & \qw \\
\wave &&&&&&&&& \\[1.5mm]
  \lstick{$\ket{v_{d[e_1]}}$} & \qw & \ctrl{1} \gategroup[2,steps=1,style={dotted,fill=cyan!20, inner xsep=1pt, inner ysep=-1pt}, background,label style={label position=above,anchor=south,yshift=-2mm}]{} & \gate{H} & \gate{X} & \gate{Z} & \gate{X} & \gate{H} & \ctrl{1} \gategroup[2,steps=1,style={dotted,fill=cyan!20, inner xsep=1pt, inner ysep=-1pt}, background,label style={label position=above,anchor=south,yshift=-2mm}]{} & \qw \\
\lstick{$\ket{v_{d[\neq]}}$} & \qw \qwbundle{k_d-1} & \targ{} & \qw & \qw & \qw & \qw & \qw & \targ{} & \qw
\end{quantikz}
}
\caption{The parallel and distributed diffuser scheme.} \label{fig:DistributedDiffuser}
\end{figure}

Let us recall the design of our parallel diffuser in Fig.~\ref{fig:ParallelDiffuser}. Since only variables $\ket{v_{j[e_1]}}$ for $j \in \{1,2, \ldots, d\}$ are selected as the representative for the diffusion process, it is natural to let $\ket{v_{j[e_1]}}$ host one node for the distributed diffusion. Fig.~\ref{fig:DistributedDiffuser} shows the design of our distributed diffuser. The critical operation is the controlled $Z$ gate (the center block in cyan color), which can be accomplished based on the proposed distributed protocol, as introduced in Section~\ref{subsec:DistributedOracle}. Except for the controlled $Z$ gate, there are other two types of operations needed to be performed distributedly:
\begin{enumerate}
    \item $\ket{v_{j[e_1]}}$ disentangles with $\ket{v_{j[\neq]}}$ for all $j \in \{1,2, \ldots, d\}$, and
    \item $\ket{v_{j[e_1]}}$ entangles back with $\ket{v_{j[\neq]}}$ for all $j \in \{1,2, \ldots, d\}$,
\end{enumerate}
\tys{Should we mention that this is the prerequisite for realizing the distributive diffuser? It might be challenging to realize this when the control and target qubits are located in different quantum computers. Or maybe we can mention that we are not concerned with the technical feasibility at this point and assumed that it can be realized readily some point in the future.}
as shown in the leftmost and rightmost cyan blocks of Fig.~\ref{fig:DistributedDiffuser}, respectively. These operations can be accomplished based on the proposed distributed protocol as well. Notice that we do not unfold the distributed protocol for each operation to be performed distributedly in Fig.~\ref{fig:DistributedDiffuser} due to the space limit. Instead, we mark those operations that can be accomplished by the proposed distributed protocol in cyan color to highlight the high-level structure of our design. Theorem~\ref{thm:DistributedDiffuserCorrect} proves the correctness of our distributed diffuser.

\begin{theorem} \label{thm:DistributedDiffuserCorrect}
Our distributed diffuser is correct.
\end{theorem}
\begin{proof}
The correctness follows from Theorem~\ref{thm:ParallelDiffuser} and Theorem~\ref{thm:ProtocalCorrectness}; the detailed proof is omitted here.
\end{proof}

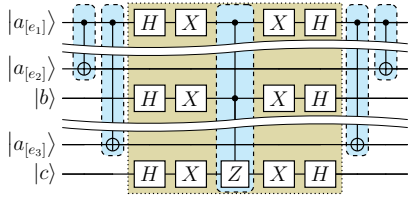
\begin{figure}[tb]
\centering
\scalebox{0.56}{
\begin{quantikz}[row sep={6mm,between origins}, column sep=2.5mm, font=\LARGE]
 \lstick{$\ket{a_{[e_1]}}$} & [1mm] \ctrl{2} \gategroup[3,steps=1,style={dashed, rounded corners,fill=cyan!20, inner xsep=0pt, inner ysep=0pt}, background,label style={label position=above,anchor=south,yshift=-0.2cm}]{} & [1mm] \ctrl{5} \gategroup[6,steps=1,style={dashed, rounded corners,fill=cyan!20, inner xsep=0pt, inner ysep=0pt}, background,label style={label position=above,anchor=south,yshift=-0.2cm}]{} & [1mm] \gate{H} \gategroup[7,steps=5,style={dotted,fill=olive!30, inner xsep=1pt, inner ysep=1pt}, background,label style={label position=above,anchor=south,yshift=-2mm}]{} & \gate{X} & [1mm] \ctrl{6} \gategroup[7,steps=1,style={dashed, rounded corners,fill=cyan!20, inner xsep=0pt, inner ysep=0pt}, background,label style={label position=above,anchor=south,yshift=-0.2cm}]{} & [1mm] \gate{X} & \gate{H} & [1mm] \ctrl{5} \gategroup[6,steps=1,style={dashed, rounded corners,fill=cyan!20, inner xsep=0pt, inner ysep=0pt}, background,label style={label position=above,anchor=south,yshift=-0.2cm}]{} & [1mm] \ctrl{2} \gategroup[3,steps=1,style={dashed, rounded corners,fill=cyan!20, inner xsep=0pt, inner ysep=0pt}, background,label style={label position=above,anchor=south,yshift=-0.2cm}]{} & [1mm] \qw \\
\wave &&&&&&&&&& \\[-1mm]
\lstick{$\ket{a_{[e_2]}}$} & \targ{} & \qw & \qw & \qw & \qw & \qw & \qw & \qw & \targ{} & \qw \\[1mm]
\lstick{\ket{b}} & \qw & \qw & \gate{H} & \gate{X} & \ctrl{1} & \gate{X} & \gate{H} & \qw & \qw & \qw \\
\wave &&&&&&&&&& \\[-1mm]
\lstick{$\ket{a_{[e_3]}}$} & \qw & \targ{} & \qw & \qw & \qw & \qw & \qw & \targ{} & \qw & \qw \\[1mm]
\lstick{\ket{c}} & \qw & \qw & \gate{H} & \gate{X} & \gate{Z} & \gate{X} & \gate{H} & \qw & \qw & \qw
\end{quantikz}
}
\caption{A parallel and distributed diffuser for $\mathcal{F}$.}\label{fig:DistributedDiffuserExample}
\end{figure}

We illustrate our approach using the running example. Fig.~\ref{fig:DistributedDiffuserExample} shows the distributed diffuser for formula $\mathcal{F}: (a) \wedge (\overline{a} \vee b) \wedge (\overline{a} \vee c)$.
Since there are three variables in $\mathcal{F}$, we need three nodes, where the first node holds $\ket{a_{[e_1]}}$, the second holds qubits $\ket{b}$ and $\ket{a_{[e_2]}}$, and the third holds qubits $\ket{c}$ and $\ket{a_{[e_3]}}$. For the controlled $Z$ gate in the diffusion process, the third node can serve as the node master in the distributed protocol. Before (resp. after) the diffusion process, $\ket{a_{[e_1]}}$ needs to disentangle (resp. entangle back) with $\ket{a_{[e_2]}}$ and $\ket{a_{[e_3]}}$. These operations can be accomplished by our distributed protocol as well. Note that our distributed protocol works only when there is one target qubit, while the structure of the disentangling/entangling operations here has one control qubit with multiple target qubits. Thus, instead of performing the disentangling/entangling in one shot, we need to perform them sequentially, e.g., $\ket{a_{[e_1]}}$ first disentangles (entangles back) with $\ket{a_{[e_2]}}$ then with $\ket{a_{[e_3]}}$, as the leftmost (rightmost) cyan blocks in Fig.~\ref{fig:DistributedDiffuserExample}. Interestingly, the order does not matter. One can easily check that different orders give the same result.

\subsection{Analysis and Simulation} \label{subsec:DistributedSATEvaluation}


As aforementioned, when our distributed diffuser tries to disentangle or entangle a variable with each of its expended variables, it has to be done sequentially using the proposed distributed protocol (Fig.~\ref{fig:mCUgate}) because the protocol works only when there is one target qubit. This introduces an overhead in time complexity. Assume that the variable $v_j \in V$ is shared by $k_j$ clauses for $j \in \{1,2,\ldots, d\}$ and $|V| = d$. Let $k_{max}$ be the maximum value among $k_j$ for all the shared variables.
The extra overhead in time complexity would be bounded by $O(k_{max})$ because the disentangling/entangling process for different variable $v_j$ is independent and can be performed in parallel.
Notice that this overhead does not exist in our parallel quantum SAT solver because the disentangling/entangling process can be done in one shot with a quantum gate in the centralized setting. 
In practice, we observe that $k_{max}$ is often much smaller than the total number of clauses in SAT solving benchmarks, so our distributed approach should still be faster than the sequential approach.

\begin{figure}[tb]
\centering
\scalebox{0.85}{
\begin{tikzpicture}
\begin{axis}  
[  
    ybar,  
    enlargelimits=0.1,  
    ylabel={Counts}, 
    xlabel={Measured Outputs ($|abc\rangle$)},  
    symbolic x coords={000, 001, 010, 011, 100, 101, 110, 111}, 
    xtick=data,  
    nodes near coords, 
    nodes near coords align={vertical},
    ymin=0,
    enlarge y limits=upper,
    height=3.2cm,
    width=9cm,
    bar width=12pt,
    ]  
\addplot coordinates {(000,251) (001,265) (010,244) (011,274) (100,258) (101,262) (110,261) (111,6377)};  
  
\end{axis}
\end{tikzpicture}
}
\caption{Simulation result of our distributed approach for formula $\mathcal{F}$.}
    \label{fig:Distributed_SAT_Result}
    \vspace{-10px}
\end{figure}
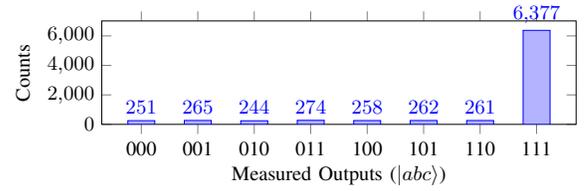

We have also implemented the distributed SAT solver for our running example $\mathcal{F}: \!(a) \wedge (\overline{a} \vee b) \wedge (\overline{a} \vee c)$ in Qiskit~\cite{GT19}. Totally, $36$ qubits are required (nine for formula $\mathcal{F}$ itself and $27$ for performing the proposed distributed quantum protocol). 
The implementation and detailed breakdown of qubits can be found in~\cite{Parallel_SAT}. Only one Grover iteration is required. Fig.~\ref{fig:Distributed_SAT_Result} shows the simulation result of performing Grover's algorithm for $8,192$ shots. The x-axis shows the measured outcome of $\ket{abc}$, while the y-axis shows the count of each outcome being measured. 
Again, the expected outcome, $\ket{111}$, has an overwhelmingly higher probability over others, which confirms the correctness of our distributed quantum SAT solving technique. 



\section{Related Work} \label{sec:RelatedWorks}

\emph{Quantum search.} 
Improving the proof search in SAT solving using quantum computing is a promising and broadly discussed direction. Barreto et al.'s method~\cite{barreto2011} adopts Shenvi's quantum random walk search algorithm~\cite{Shenvi2003} in a local search setting and applies it to 3-SAT --- a specialized SAT solving with 3 variables in each clause. Their method enables parallel simulation of the quantum SAT solving algorithm, though it is different from our notion of performing and coordinating multiple quantum SAT solving instances in parallel. 

Another prominent example is to use Grover's algorithm to search for a satisfiable truth assignment for Boolean variables~\cite{fernandes2019}. 

\emph{Hybrid methods.} 
A straightforward application of Grover's algorithm in SAT solving requires a large number of qubits. Consequently, several \emph{hybrid} approaches are proposed to reduce the number of qubits by combining quantum computing with classical computing algorithms. For example, quantum cooperative search replaces some qubits with classical bits and solves the classical bits using traditional SAT solving~\cite{Cheng2007}. Zhang et al.'s approach optimize the data structures in SAT solving to take advantage of Grover's algorithm and DPLL~\cite{zhang2020}. Another venue is to focus on a parameterized area of the search space and then Grover's search~\cite{Varmantchaonala2023}. These hybrid approaches achieved varied theoretical improvements in the time complexity of SAT solving.

\emph{Quantum heuristics.}
Quantum walk~\cite{Childs2003} may also be applied in heuristics that improve SAT solving. Campos et al.~\cite{campos2021} presented such an algorithm for solving $k$-SAT, where each clause has exactly $k$ variables. Their approach leverages continuous time quantum walk over a hypercube graph with potential barriers. Their construction of the problem exploits the properties of quantum tunnelling to obtain the possibility of getting out of local minima. Their simulation shows a reasonable successful rate, though heuristic methods may not guarantee that a solution is found. Thus, this kind of research has a different goal from ours. Similarly, research on classical algorithms for quantum SAT solving~\cite{aldi2021} is also in a different vein.

\emph{Quantum annealing.} Some of the above techniques may be deemed quantum optimizers. Quantum annealers~\cite{apolloni1990} are another widely used optimization technique that minimizes objective functions over discrete variables using quantum fluctuation. Bian et al.'s method~\cite{Bian2017} encodes SAT solving into a quadratic unconstrained binary optimization (QUBO) problem and applies quantum annealing to solve it.

\emph{Applications.}
Quantum SAT solving has found numerous applications. For instance, Quantum SAT solving may be applied to speed up integer factorization. Mosca et al.~\cite{mosca2020} showed how to design SAT circuits for finding smooth numbers, which is an essential step in Number Field Sieve (NFS) --- the best-known classical solution. Assuming that there is a quantum SAT solver that performs better than classical solvers, their method would lead to a factorization method that outperforms NFS. The maximum satisfiability (MAX-SAT) problem asks for the maximum number of clauses that are satisfiable in a conjunctive normal form. Alasow and Perkowski~\cite{Alasow2022} apply Grover's search with a customized oracle to perform SAT solving, which also leads to an efficient solution to MAX-SAT.


Qiu et al. proposed a distributed Grover's algorithm~\cite{QLX22}, which decomposes the original SAT formula into a set of $2^k$ subformulas (obtained by instantiating $k$ Boolean variables). Each of the $2^k$ subformulas is then solved by one quantum computer running Grover's algorithm, and the final solution depends on the subsolutions to the subformulas. Their ``divide and conquer'' strategy does not utilize any quantum characteristics, while ours utilizes quantum teleportation.
\tys{the protocol mentioned in the distributive setting is not a conventional quantum teleportation, would it be better to replace as "quantum nonlocality"?}

\section{Conclusion and Future Work} \label{sec:Conclusion}

This work is the first to propose a parallel quantum SAT solver 
using entanglement.
Compared to the sequential quantum SAT solver, our parallel solver reduces the time complexity of each Grover iteration from linear time $O(m)$ to constant time $O(1)$ by using more qubits. 
To scale to complex problems, we also propose the first distributed quantum SAT solver using quantum teleportation such that the total qubits required are shared and distributed among a set of quantum computers (nodes), and the quantum SAT solving is accomplished collaboratively by all the nodes. We prove the correctness of our methods. They are also evaluated in simulations via Qiskit, and the results are correct. 
In the future, we plan to extend our parallel and distributed quantum SAT solvers to handle satisfiability modulo theories (SMT) problems.


\bibliographystyle{plain}
\bibliography{reference}

\end{document}